\documentclass[pra,showpacs,twocolumn]{revtex4}
\usepackage{bm}
\usepackage{amsmath}
\usepackage{amssymb}
\usepackage{graphicx}

\newcommand{\eqb}{\begin{eqnarray}}
\newcommand{\eqe}{\end{eqnarray}}
\newcommand{\rd}{\mathrm{d}}
\newcommand{\bx}{{\bf x}}
\newcommand{\bk}{{\bf k}}

\newcommand{\bQ}{{\bf Q}}

\newcommand{\pd}{\partial}

\begin{document}

\title{Nonlinear intraband tunneling of BEC in a cubic three-dimensional lattice}
\author{V. S. Shchesnovich$^{1}$ and V. V. Konotop$^2$}
\affiliation{
$^1$ Instituto de F\'{\i}sica - Universidade Federal de Alagoas, Macei\'o AL
57072-970, Brazil
\\
$^2$ Centro de F\'isica Te\'orica e Computacional, Universidade de Lisboa, Complexo
Interdisciplinar, Avenida Professor Gama Pinto 2, Lisboa 1649-003, Portugal;
Departamento de F\'isica, Faculdade de Ci\^encias, Universidade de Lisboa, Campo
Grande, Ed. C8, Piso 6, Lisboa 1749-016, Portugal
}


\begin{abstract}
The intra-band  tunneling of a Bose-Einstein condensate between three degenerate
high-symmetry $X$-points of the Brillouin zone of a cubic optical lattice is
studied in the quantum regime by reduction to a three-mode model. The mean-field
approximation of the deduced model is described. Compared to the previously
reported two-dimensional (2D) case [Phys. Rev. A \textbf{75,} 063628 (2007)], which
is reducible to the two-mode model, in the case under consideration there exist a
number of  new stable stationary atomic distributions between the $X$-points and a
new critical lattice parameter. The quantum collapses and revivals of the atomic
population dynamics are absent for the experimentally realizable time span. The 2D
stationary configurations, embedded into the 3D lattice, turn out to be always
unstable, while existence of a stable 1D distribution, where all atoms populate
only one $X$-state, may serve as a starting point in the experimental study of the
nonlinear tunneling in the 3D  lattice.
\end{abstract}

 \maketitle

\section{Introduction}

In recent  paper~\cite{SK} it has been shown that exploring the nonlinear tunneling
of a Bose-Einstein condensate (BEC) loaded in a two-dimensional (2D) optical
lattice allows for theoretical and experimental study of diversity of fundamental
issues of the nonlinear physics. Among them we mention the validity of the
mean-field approximation (considered previously in Ref. \cite{meanfiled}), which
was used in the study of the nonlinear tunneling \cite{BKK,BKKS}, the accuracy of
the semi-classical (i.e. WKB) approximation, where the inverse number of atoms
$1/N$ plays the role of the effective Planck constant (similar to the ideas
reported in Ref. \cite{VA} for a coupled two-mode model and in Ref. \cite{KonKev}
for the Bohr-Sommerfeld quantization rule), and the macroscopic manifestation of
the quantum collapses and revivals -- a pure quantum effect related to the discrete
nature of quantum spectra~\cite{revivals}. The approach developed in Ref.~\cite{SK}
was based on the energy degeneracy due to the rotational, more specifically C$_4$,
symmetry of the lattice, where  the modulationally stable Bloch states at the
high-symmetry $X$-points were used (for either attractive BEC loaded in the first
lowest band or repulsive BEC loaded in the second band). The modulational stability
allows one to reduce the  mean-field description of spatially inhomogeneous matter
waves to the effective two-mode model describing the populations of the resonant
states (see also~\cite{BK} and the references therein). It was shown in
Ref.~\cite{BKKS} by direct numerical simulations that the  two-mode model gives a
remarkably well description of the dynamics for  a relatively long interval of
time.

Since optical lattices are routinely available in all dimensions (see, for
instance, the recent review \cite{3Dlattice}), an intriguing possibility is to
explore the dynamics similar that reported in Ref.~\cite{SK} but in the 3D case.
Considering a cubic lattice one expects the dynamics to be much richer as compared
to the 2D case, because the distinct X-points of the same band are now three-fold
degenerate. Concentrating on the modulationally stable case, the only situation
considered in the present paper, one would expect that the stable distributions are
very different from those in the 2D configuration, moreover, the latter (embedded
in the 3D lattice) is found to be unstable. The most significant feature of the
problem at hand is that it reduces to an effective \textit{three-mode model}, whose
dynamics is described by a Hamiltonian with two degrees of freedom, due to the
constraint imposed by  conservation of the number of atoms. Taking into account
that such dynamics in the vicinity of a stable point is, generally speaking,
characterized by two frequencies, as well as the facts that in most of the spectrum
the energy level spacing in the quantum model scales at least as $N^{-2}$ (except,
for example, the local bound states close to the semiclassical stationary points)
and that the number of atoms used in BEC experiments is large, one can expect that
the phenomenon of quantum collapses and revivals in the respective quantum system
is significantly affected (and even suppressed completely). At the same time new
features can be expected due to the fact that the classical motion now can be
either regular (in the vicinity of the stationary points) or chaotic, what will
naturally affect the underlying quantum evolution.

Study of the phenomena mentioned above with special attention payed to the
correspondence between quantum and semi-classical dynamics (like in the 2D case,
the semi-classical  dynamics will be obtained in the mean-field approach)
constitutes the main goal of the present paper.

More specifically, we start by  introducing in Sec.~\ref{model} the quantum model
and discussing its validity, physical parameters and the time span achievable in
possible experiments. The mean-field limit is derived in Sec.~\ref{quasiclass} by
making use of the WKB approximation for the quantum model rewritten as a
Schr\"odinger equation for an effective quantum particle. We study the stationary
points of the mean-field dynamics and their stability by  locally diagonalizing the
Hamiltonian. In Sec.~\ref{numerics} we compare the numerical simulations of the
quantum model to those of the  mean-field approach. Finally, in
Sec.~\ref{conclusion} we summarize the results.

\section{The reduced quantum  model}
\label{model}

\subsection{The reduced Hamiltonian}

Let us start with the Hamiltonian of a BEC in an optical lattice
\eqb
H = \int\limits_\mathcal{V}\rd^3\bx
\psi^\dag(\bx)\left(-\frac{\hbar^2}{2m}\nabla^2+V(\bx)\right)\psi(\bx)
\nonumber \\
+\frac{g}{2}\int\limits_\mathcal{V}\rd^3\bx
\psi^\dag(\bx)\psi^\dag(\bx)\psi(\bx)\psi(\bx),
\label{EQ1}
\eqe
where  ${\bf x}\in\mathbb{R}^3$, $V(\bx)$ is the cubic optical lattice potential,
$\mathcal{V}=Mv_0$ is the total volume of the lattice consisting of $M$ cells each
one of the volume $v_0$, $g$ is the interaction coefficient, and $\psi^\dag({\bf
x})$ and $\psi({\bf x})$ are the creation and annihilation field operators.
Introducing the Bloch waves $\varphi_{n\bk}(\bx)$ through the standard eigenvalue
problem
\[
\left(-\frac{\hbar^2}{2m}\nabla^2+V(\bx)\right)\varphi_{n\bk}=E_{n\bk}\varphi_{n\bk},
\]
where $n$ is a number of the zone and $\bk$ is the wave-vector in the first
Brillouin zone (BZ), we expand
\eqb
\psi(\bx) = \sum_{n,\bk}\varphi_{n,\bk}(\bx)b_{n,\bk},
\label{EQ2}
\eqe
where the creation and annihilation operators satisfy the usual commutation
relations $[b_{n\bk},b_{n^\prime\bk^\prime}]=0$ and
$[b_{n\bk},b^\dag_{n^\prime\bk^\prime}] =
\delta_{nn^\prime}\delta_{\bk\bk^\prime}$. The expansion (\ref{EQ2}) allows one to
rewrite the Hamiltonian (\ref{EQ1}) in the form
\eqb
&&H = \sum_{n,\bk}E_{n\bk}b^\dag_{n\bk} b_{n\bk} +
\nonumber \\
&& \sum_{\bk_1,...,\bk_4 \atop n_1,...,n_4}
\chi^{n_1n_2n_3n_4}_{\bk_1\bk_2\bk_3\bk_4}
\delta_{\bk_1+\bk_2-\bk_3-\bk_4,\bQ}b^\dag_{n_1\bk_1} b^\dag_{n_2\bk_2}
b_{n_3\bk_3}b_{n_4\bk_4},
\nonumber \\
\label{EQ3}\eqe
where $\bQ$ is an arbitrary vector of the reciprocal lattice and
\eqb
\chi^{n_1n_2n_3n_4}_{\bk_1\bk_2\bk_3\bk_4} = \frac{g}{2}
\int\limits_{\mathcal{V}}\rd^2\bx
\varphi^*_{n_1\bk_1}\varphi^*_{n_2\bk_2}\varphi_{n_3\bk_3}\varphi_{n_4\bk_4}
\label{EQ4}
\eqe
(hereafter the asterisk stands for the complex conjugation).

By analogy with the 2D case \cite{SK}, after the symmetry of the lattice is fixed
the nonlinear tunneling phenomenon in 3D does not depend on the particular shape of
the potential,  as for instance on its separability: the only lattice parameter
$\Lambda$ entering the model, see Eq. (\ref{Heff}) below, characterizes the
effective lattice depth rather than its geometric properties. Therefore we
concentrate on the simplest case of a separable cubic lattice:
\eqb
V = sE_r[\cos(x/\rd) + \cos(y/\rd)+\cos(z/\rd)],
\label{LATT}
\eqe
where  $\rd$ is the lattice period and the lattice depth $s$ is measured in the units of
the recoil energy $E_r = \hbar^2\pi^2/(2md^2)$. The respective BZ is given by
$[-\frac{\pi}{\rd},\frac{\pi}{\rd}]\times[-\frac{\pi}{\rd},\frac{\pi}{\rd}]\times[-\frac{\pi}{\rd},\frac{\pi}{\rd}]$.

We consider the three distinct $X$-points, $X_1 = (\frac{\pi}{d},0,0)$, $X_2 =
(0,\frac{\pi}{d},0)$ and $X_3 = (0,0,\frac{\pi}{d})$, degenerate in the Bloch
energy, which pertain to the modulationally stable Bloch band, and assume that only
these points are significantly populated by the BEC atoms at $t=0$. Then, repeating
the arguments of the 2D case \cite{SK} it can be shown that for small $g$ (see also
below) the energy and quasi-momentum conservation laws allow one to discard the
transitions, due to the  scattering of BEC atoms, to all other points except the
transitions between the three degenerate $X$-points. We arrive at the three-mode
approximation:
\eqb
\psi(\bx) = \varphi_1(\bx)b_1 + \varphi_2(\bx)b_2 +\varphi_3(\bx)b_3,
\label{EQ5}\eqe
where the Bloch function $\varphi_j(\bx)$ and the operator $b_j$ correspond to the
$X_j$-point (here we use the simplified-index notations for the populated states).
Substitution of this expression into Eq. (\ref{EQ1}) results in the approximate
Hamiltonian
\begin{eqnarray}
\label{EQ6}
 H_{X}&=&\sum_{j=1}^{3}\left(E_X b^\dag_j b_j+\chi_1 b_j^\dag b_j^\dag b_jb_j\right)
\nonumber \\
&+&\chi_2 \left(4\sum_{j<k}b_j^\dag b_k^\dag b_jb_k + \sum_{j\ne k}(b_j^\dag)^2
(b_k)^2\right)
\end{eqnarray}
where $E_X$ is the Bloch energy at the $X$-point and the only nozero coefficients (\ref{EQ4}) are given by
\eqb
\chi_1= \frac{g}{2} \int\limits_{\mathcal{V}}\rd^3\bx\, |\varphi_1|^4, \quad
\chi_2= \frac{g}{2} \int\limits_{\mathcal{V}}\rd^3\bx\, |\varphi_1|^2|\varphi_2|^2.
\label{EQ7}
\eqe
The Hamiltonian (\ref{EQ6}) commutes with the total number of atoms $N$ in the $X$-points
\begin{eqnarray}
\label{N_cons}
\sum_{j=1}^3n_j=N, \quad n_j=b_j^\dag b_j
\end{eqnarray}
 which reflects the approximation made.


The details of the derivation can be found in Ref. \cite{SK}. Let us, however,
present an alternative way to arrive at the Hamiltonian (\ref{EQ6}). First we note
that if initially only the $X$-points of the same Bloch band are populated, the
rates of the quantum transitions to other points (within the same band and to other
Bloch bands) due to the $s$-wave atomic scattering (i.e. the nonlinearity of BEC),
treated as a perturbation, are defined, according to the Fermi golden rule, by the
energy conservation between the initial and final states within the linear model
and are proportional to the density of the states at the particular point to which
the transition takes place. In our particular case, an additional rule on the
transitions, satisfied by the nonlinearity of BEC, is that the sum of the
quasi-momenta   of the two atoms before the scattering and after that can differ
only by some reciprocal lattice vector (we consider the Bloch waves as the  basis).
Recalling that the density of states is large only  on the boundary of the
Brillouin zone (it diverges in the limit of an infinite lattice) and only the
$X$-points   of the same Bloch band on the boundary have the same energy and
satisfy the quasi-momentum conservation (in the above sense), all transitions
except those to the $X$-points of the same band can be neglected. The same
conclusion is also derived within the mean-field approach. Indeed, the resonant
four-wave processes are determined by the phase-matching conditions (equivalent to
the energy and the quasi-momentum conservation) and the population of the
respective points, thus unpopulated  points (or entire bands) which do not satisfy
the matching conditions  do not make any contribution to the resonant processes.

In this set of arguments, the  transitions to other (resonant) points of the same
Bloch band  are neglected due to either the negligibly low density of states
compared to that at the boundary points or due to the quasi-momentum
non-conservation. On the other hand, the transitions to the boundary points of
other Bloch bands are neglected due to the energy conservation, i.e. under the
condition that the nonlinearity of BEC is much smaller than the band gap at the
boundary of the Brillouin zone. In the shallow lattice limit $s\ll 1$ ($\Lambda$
close to 1), for instance, the condition of a relatively large gap requires that
$\chi_1 N \sim gN/{\cal V}\ll E_r$.

On the other hand, we use the expansion over the Bloch-wave basis [see Eq.
(\ref{EQ2})],  what means that a few-mode approximation implies concentration of
particles in the respective states. This imposes a constraint on the potential
depth. Indeed, two body interactions result in nonzero spectral width of a Bloch
state, which in our case can be estimated as $\omega_{NL} =
\frac{4\chi_{1}N}{\hbar} $. In order to be able to neglect the effect of the
spectral width on the dynamics (what is done in the present paper) one has to
require it to be much less than the spectral width of the lowest band (the most
narrow band in the general situation). The latter is typically of order of the
recoil energy $E_r$. Thus we require $\omega_{NL}\ll E_r/\hbar $ (notice that that
this is the limit in some sense opposite to the standard conditions of
applicability of the Bose-Hubbard model, where due to relatively large amplitude of
the periodic potential and, consequently, strong on-site localization of the wave
functions, the expansion over Wannier functions is more appropriate). In order to
get an idea about the physical range of the parameters, let us estimate the
frequency $\omega_{NL}$. The coefficient $\chi_1 = \frac{g}{2}\int\rd
x^3|\varphi|^4 \sim \frac{g}{2\mathcal{V}}$, hence  we can estimate
\begin{eqnarray}
\omega_{NL} \sim \frac{8\pi\hbar a_s}{m} \frac{N}{\mathcal{V}}= \frac
1\gamma\frac{a_s}{m} \frac{N}{\mathcal{V}}, \label{TNL}
\\
\gamma = 3.78 \times
 10^{32}(J\cdot s)^{-1}. \nonumber
\end{eqnarray}
For instance, for ${}^{87}$Rb we have $m = 1.44\times 10^{-25}kg$ and $a_s = 5.1
nm$ thus for a lattice with $M=20^3$ cells with the lattice constant $d=1\mu m$
($v_0=10^{-12}$cm$^3$) and  $N = 1000$ we get $\omega_{NL}\sim 12.5 Hz$.   On the
other hand assuming the potential depth to be equal to the recoil energy we obtain
the width of the first lowest band to be $0.59\cdot E_r/\hbar\approx 590$ Hz, and
thus $\hbar\omega_{NL}/E_r\approx 0.021$. By reducing the potential depth this
relation can be further improved.

In  this context it is also relevant to mention, that the dimensional time $\tau$
is measured in the units $1/\omega_{NL}$, which for the above example is
approximately $0.08$ s. Taking into account that the characteristic lifetime of a
condensate can reach $10 s$, we conclude that for typical  experiments the
characteristic time is about $100$ dimensionless units. This time can be
significantly enlarged (i.e. making observation of the reported effect much easier)
by using lighter, say lithium, atoms and/or with a larger $s$-wave scattering
length, achievable by Feshbach resonance.

\subsection{The dynamical equations}

It is convenient to use the dimensionless time $\tau = \omega_{NL}t $, subtract
from the Hamiltonian (\ref{EQ6}) the constant term $H_0 = (E_X-\chi_{1})N$ [here we
use Eq. (\ref{N_cons})] and normalize the result by dividing by $N^{2}$, which
allows for the transition to the mean-field limit $N\to\infty$,  since the
resulting  Hamiltonian is written in the population densities $n_j/N$. This results
in the Schr\"{o}dinger equation
\begin{equation}
\frac{i}{N}\partial_\tau|\Psi\rangle =\Hat{H}|\Psi\rangle,
\label{SCHR}
\end{equation}
with the Hamiltonian
\begin{eqnarray}
\Hat{H}
 =\frac{1}{N^{2}}\left(\frac{1}{4}\sum_{j=1}^3n_j^2 +\Lambda\sum_{j<k}n_jn_k
+\frac{\Lambda}{4}\sum_{j\ne k}(b^\dag_j)^2b_k^2\right)\!,
\label{Heff}
\end{eqnarray}
where $\Lambda= \chi_2/\chi_1$. Equation (\ref{Heff}) may be interpreted as a
Schr\"odinger equation for a single quantum particle (see also Eq. (\ref{EQ12})
below).

Denoting by $k_j$ the number of atoms populating the  X$_j$-point (such that
$k_1+k_2+k_3=N$) one can expand the wave function $|\Psi\rangle$ over the Fock
basis \mbox{$|k_1,k_2,N-k_1-k_2\rangle \equiv |k_1\rangle |k_2\rangle
|N-k_1-k_2\rangle$}:
\begin{eqnarray}
    |\Psi\rangle=\sum_{k_1=0}^{N}\sum_{k_2=0}^{N-k_1}C_{k_1,k_2}(t)|k_1,k_2,N-k_1-k_2\rangle,
\label{EQ8}
\end{eqnarray}
where the expansion coefficients obey the normalization condition
\begin{eqnarray}
 \sum_{k_1=0}^{N}\sum_{k_2=0}^{N-k_1}|C_{k_1,k_2}(t)|^2=1.
\label{EQ9}
\end{eqnarray}
Now Eq. (\ref{Heff}) can be cast in the  form:
\begin{widetext}
\begin{eqnarray}
    \frac{i}{N}\frac{dC_{k_1,k_2}}{d\tau}&=&\frac{1}{4}a_{k_1,k_2}C_{k_1,k_2}+\frac{\Lambda}{4}\left(
    b_{k_1-1,k_2}  C_{k_1-2,k_2}
        +b_{k_1+1,k_2}  C_{k_1+2,k_2} +b_{k_2-1,k_1}  C_{k_1,k_2-2}+b_{k_2+1,k_1}  C_{k_1,k_2+2}
    \right.
     \nonumber \\
    & +&\left. d_{k_1-1,k_2+1}  C_{k_1-2,k_2+2}+d_{k_1+1,k_2-1} C_{k_1+2,k_2-2}
    \right),
\label{EQ10}
\end{eqnarray}
where
\begin{eqnarray*}
&&a_{k_1,k_2}= 1 + 2(2\Lambda -1)N^{-2}\left[(k_1 +k_2)(N - k_1 -k_2)
+k_1k_2\right],
 \\
&&b_{k_1,k_2}=N^{-2}\left\{k_1(1+k_1)(N-k_1-k_2+1)(N-k_1-k_2)\right\}^{1/2},
  \\
&&d_{k_1,k_2}=N^{-2}\left\{k_1(1+k_1)k_2(1+k_2)\right\}^{1/2}.
\label{EQ11}\end{eqnarray*}
\end{widetext}
We notice that the coefficients are defined only for $0\le k_1+k_2\le N$, which is
the lower left triangular part of the corresponding matrix representation and the
coefficient $b_{k_1,k_2}$ is not symmetric with respect to the exchange of the
indexes.

The nonlinearity, though being responsible for the very existence of the intraband
tunneling, only defines the time scale and it is the lattice parameter $\Lambda$
which enters the Hamiltonian in Eq.  (\ref{Heff}) ($\Lambda$ is a ratio of two
integrals of the Bloch waves which are defined solely by the lattice).

We conclude this section with the estimate for the energy range:
\begin{eqnarray}
&&\frac{1+2\Lambda}{12} +\frac{\Lambda}{2N} \le \langle \Hat{H}\rangle \le
\frac{1+2\Lambda}{4} -\frac{\Lambda}{2N}, \; \Lambda \le \frac{1}{4}\nonumber\\
&&\frac{1-2\Lambda}{4} +\frac{\Lambda}{2N} \le \langle \Hat{H}\rangle \le
\frac{1+2\Lambda}{4} -\frac{\Lambda}{2N}, \; \Lambda > \frac{1}{4}, \nonumber\\
&&
\label{EST}
\end{eqnarray}
(see Appendix~\ref{energy}) important for the numerical simulations. Since the
Hamiltonian (\ref{Heff}) is bounded from above and from below it follows that the
energy spacing for the quantum particle satisfying equation (\ref{SCHR}) is, for
the most of the spectrum, on the order of $\delta E \sim N^{-2}$: for a given
number of atoms $N$ the dimension of the Hilbert space is $(N+1)(N+2)/2$ i.e.
$\approx N^2/2$ in the limit $N\ll 1$ (in the 2D case, considered in
Ref.~\cite{SK}, the dimension of the respective Hilbert space was $N+1$ and
respectively the energy distance between adjacent energy levels was determined by
the factor $N^{-1}$).
\medskip


\section{The semi-classical approximation}
\label{quasiclass}

\subsection{The governing dynamical model }

The semi-classical approach employed here is similar to that of Ref.~\cite{SK}. We
define $h=2/N$, $x_{1,2}= k_{1,2}/N$. Then, assuming existence of a regular
function $\psi(x_1,x_2) \equiv \frac{N+1}{2}C_{k_1,k_2}$, Eq. (\ref{EQ10}) is cast
as
\begin{widetext}
\eqb
&&ih\pd_\tau \psi =\frac{1}{2}a(x_1,x_2)\psi +
\frac{\Lambda}{2}\biggl\{b_h\left(x_1-\frac{h}{2},x_2\right)e^{-i\hat{p}_1} +
b_h\left(x_1+\frac{h}{2},x_2\right)e^{i\hat{p}_1}+b_h\left(x_2-\frac{h}{2},x_1\right)e^{-i\hat{p}_2}
\nonumber \\
&& + b_h\left(x_2+\frac{h}{2},x_1\right)e^{i\hat{p}_2} +
d_h\left(x_1-\frac{h}{2},x_2+\frac{h}{2}\right)e^{i(\hat{p}_2-\hat{p}_1)}
+d_h\left(x_1+\frac{h}{2},x_2-\frac{h}{2}\right)e^{i(\hat{p}_1-\hat{p}_2)}\biggr\}\psi,
\nonumber \\
&& \label{EQ12}
\eqe
where  $\hat{p}_j = -ih\pd_{x_j}$ and we have introduced the functions, defined on
the triangular domain $0\le x_1+x_2 \le 1$:
\begin{eqnarray*}
&& a(x_1,x_2) =  1 +
2(2\Lambda-1)\left[b_0(x_1,x_2)+b_0(x_2,x_1)+d_0(x_1,x_2)\right],
\nonumber\\
&& b_h(x_1,x_2) =
\left[x_1\left(x_1+\frac{h}{2}\right)\left(1-x_1-x_2\right)\left(1-x_1-x_2+\frac{h}{2}\right)\right]^{1/2},
\nonumber\\
&&
d_h(x_1,x_2)
=\left[x_1\left(x_1+\frac{h}{2}\right)x_2\left(x_2+\frac{h}{2}\right)\right]^{1/2}.
\label{EQ13}
\end{eqnarray*}
\end{widetext}
We have $[\hat{p_j},x_k] = -ih\delta_{j,k}$, i.e. the usual canonical commutator of
the momenta and coordinates. Evidently, the r\^ole played by $h$ is of the
effective Planck constant. The semi-classical dynamics corresponds to the limit
$h\to 0$, i.e. when the number of BEC atoms $N\to\infty$. It is important to recall
that the characteristic time $t$ of the evolution scales as $(\chi_1N)^{-1}\tau$,
hence the quantity $\chi_1N$ must be kept fixed. These two conditions taken
together constitute the usual mean-field limit for BEC. It is important to mention
here that the coefficients $\chi_{1,2}$ stay bounded, as it follows form the
definition (\ref{EQ7}) and the normalization of the wave function, implying
$\int|\varphi_j|^2|\varphi_k|^2\sim 1/{\cal V}$. The quantity $\chi_1N\sim gN/{\cal
V}$ giving the scale of the nonlinear time $T_{NL}=2\pi/\omega_{NL}$ (see Eq.
(\ref{TNL})) is constant also in the thermodynamic limit defined as $N\to\infty$ at
a constant density.

The limit $h\to0$, if it exists, corresponds to the continuous limit of the
discrete equation (\ref{EQ12}) (in this respect, it is similar to the WKB approach
used for the discrete three-term relation, see Ref. \cite{braun} for details).

In order to derive the classical  equation corresponding to the limit $h\to0$ we
set $\psi(x_1,x_2,\tau) = e^{iS(x_1,x_2,\tau,h)/h}$  for a complex action
$S(x_1,x_2,\tau,h)$ viewed as a series $S = S^{(0)} +hS^{(1)} + O(h^2)$. Assuming
the action $S^{(0)}(x_1,x_2,\tau)$ be differentiable function we get the
Hamilton-Jacobi equation  for the classical action $S^{(cl)}(x_1,x_2,\tau) =
S^{(0)}(x_1,x_2,\tau) -\tau/2$:
\begin{widetext}
\begin{eqnarray}
-S^{(cl)}_\tau &=&  b_0(x_1,x_2)\left[\Lambda\cos\left(S^{(cl)}_{x_1}\right)
+2\Lambda-1\right]+ b_0(x_2,x_1)\left[\Lambda\cos\left(S^{(cl)}_{x_2}\right)
+2\Lambda-1\right]
\nonumber\\
&&+ d_0(x_1,x_2)\left[\Lambda\cos\left(S^{(cl)}_{x_1}-S^{(cl)}_{x_2}\right)
+2\Lambda-1\right]\equiv {\mathcal
H}\left(S^{(cl)}_{x_1},S^{(cl)}_{x_2},x_1,x_2\right),
\label{EQ14}\end{eqnarray}
\end{widetext}
where we have introduced the classical Hamiltonian $\mathcal{H}$.  The
quasi-classical dynamics is sometimes more conveniently described in terms of
variables $z_j = 1-2x_j $ and $\phi_j = \tilde{S}^{(cl)}_{x_j}=p$, where $x_j$ and
$p_j$ are the classical limits of the corresponding quantum variables. The Poisson
brackets of the respective classical variables read
\begin{eqnarray}
&&\{\phi_j,z_k\} = \lim_{h\to0}\frac{i}{h}[\hat{p}_j,1-2x_k] = -2\delta_{j,k},
\nonumber\\
&&\{\phi_j,\phi_k\} =\{z_j,z_k\} = 0.
\label{COMM}\end{eqnarray}

The classical variables can be associated with  the quantum averages  by the
following correspondence:
\begin{eqnarray}
z_j &=& 1 - \frac{2}{N}\langle b_j^\dag b_j\rangle = 1
-\frac{2}{N}\sum_{k_1=0}^N\sum_{k_2=0}^{N-k_1}k_j|C_{k_1,k_2}|^2,\quad
\label{zj}
\\
\phi_1 &=& \mathrm{arg}\langle(b_1^\dag)^2 b_3^2\rangle \nonumber\\
&=&
\mathrm{arg}\left\{\sum_{k_1=0}^N\sum_{k_2=0}^{N-k_1}C^*_{k_1,k_2}b_{k_1+1,k_2}C_{k_1+2,k_2}\right\},
\label{phi1}\end{eqnarray}
\begin{eqnarray}
\phi_2 &=& \mathrm{arg}\langle(b_2^\dag)^2 b_3^2\rangle \nonumber\\
&=&
\mathrm{arg}\left\{\sum_{k_1=0}^N\sum_{k_2=0}^{N-k_1}C^*_{k_1,k_2}b_{k_2+1,k_1}C_{k_1,k_2+2}\right\}.
\label{phi2}\end{eqnarray}
The first equalities in these formulae can be most easily established by
replacement of the boson operators by $c$-numbers:
$b_j^\dag\to\sqrt{N}(b^{(cl)})^*$ and $b_j\to \sqrt{N} b^{(cl)}$. The phase
difference  $\phi_j$ in Eqs. (\ref{phi1})-(\ref{phi2}) is not defined if $\langle
b_j^\dag b_j\rangle = 0$, i.e. $z_j=1$ (the function ``arg'' in Eq. (\ref{phi1}) or
(\ref{phi2}) is applied to  zero: $C_{k^\prime_1,k^\prime_2}=0$ for $k^\prime_j\ge
1$). The two phases are not defined also for $z_1+z_2 = 0$, i.e. $\langle b_3^\dag
b_3\rangle =0$, since in this case $b_3|\Psi\rangle = 0$. In these cases the phases
can be determined by taking the averages of the boson operators corresponding to
non-zero average populations, i.e. in the semi-classical limit instead of
$(b^*_jb_k)^2$ one just takes the phase of the squared nonzero amplitude $b_k^2$ or
$(b^*_j)^2$.

The classical Hamiltonian, recovered from the Hamilton-Jacobi equation
(\ref{EQ14}), reads
\begin{eqnarray}
    &&\mathcal{H}=\frac{1}{4}(1-z_1)(z_1+z_2)(\Lambda\cos\phi_1+2\Lambda-1)
    \nonumber \\
      &&+ \frac{1}{4}(1-z_2)(z_1+z_2)(\Lambda\cos\phi_2+2\Lambda-1)
    \nonumber \\
     &&+\frac{1}{4}(1-z_1)(1-z_2)(\Lambda\cos(\phi_1-\phi_2)+2\Lambda-1).
\label{EQ15}
\end{eqnarray}
As a result of Eq. (\ref{COMM}) the mean-field equations of motion acquire the form
($j=1,2$)
\eqb
\frac {dz_j}{d\tau}=-2\frac{\partial \mathcal{H}}{\partial \phi_j},\qquad \frac
{d\phi_j}{d\tau}=2\frac{\partial \mathcal{H}}{\partial z_j}.
\label{EQ16}
\eqe
Explicitly they read
\begin{widetext}
\begin{subequations}
\label{mean-field}
\begin{eqnarray}
\dot{z}_1&=&\frac{\Lambda}{2}(1-z_1)[(z_1+z_2)\sin
\phi_1 +(1-z_2)\sin(\phi_1-\phi_2)],
\\
\dot{z}_2&=&\frac{\Lambda}{2}(1-z_2)\left[(z_1+z_2)\sin
\phi_2+(1-z_1)\sin(\phi_2-\phi_1)\right],
\\
\dot{\phi}_1&=&\frac{1}{2}(1-2z_1-z_2)(\Lambda\cos\phi_1+2\Lambda-1)
+\frac{\Lambda}{2}(1-z_2)[\cos\phi_2-\cos(\phi_1-\phi_2)],
\\
\dot{\phi}_2&=&\frac{1}{2}(1-2z_2-z_1)(\Lambda\cos\phi_2+2\Lambda-1)
+\frac{\Lambda}{2}(1-z_1)[\cos\phi_1-\cos(\phi_1-\phi_2)].
\end{eqnarray}
\end{subequations}
\end{widetext}

\subsection{Stationary points}

Either from the point of view of dynamics governed by Eq. (\ref{mean-field}) or
from the point of view of practical applications the most relevant first step in
studying the mean-field dynamics is the investigation of stationary points
$P_j=\{z_{1,j}^{(st)},z_{2,j}^{(st)},\phi_{1,j}^{(st)},\phi_{2,j}^{(st)} \}$  and
of their stability. We emphasize that now the stability is understood in the
classical mechanics sense, unlike the modulational instability of the Bloch states
mentioned in the Introduction and resulting in developing of the spatial
structures~\cite{BKK,BKKS,BK}.

We will  use the notations  $\zeta_\alpha= z_\alpha-z_{\alpha,j}^{(st)}$ and
$\varphi_\alpha=\phi_\alpha-\phi_{\alpha,j}^{(st)}$ for local coordinates
describing small deviations from the stationary solutions. The relation between the
populations $x_j$ and dynamical variables $z$ reads
\begin{eqnarray}
\label{x-z}
    (x_1,x_2,x_3)=\frac 12 (1-z_1,1-z_2,z_1+z_2).
\end{eqnarray}
It is convenient to separate {\em internal} stationary points, i.e. the ones for
which all three $X$-points are populated from the {\em boundary} stationary points
for which either one or two $X$-points have zero population. The boundary
stationary points correspond to the effectively low-dimensional (2D or 1D)
distributions of atoms in the 3D lattice.

\subsubsection{Internal stationary points}

1. The first internal stationary point is given by $P_1=\left\{ \frac 13,\frac
13,0,0\right\}$ and describes equally populated  X-points with zero phases (phase
differences). The Hamiltonian in the vicinity of $P_1$ reads
\begin{eqnarray}
\mathcal{H}_{P_1} &=& \frac{1}{4}(1-3\Lambda)(\zeta_1^2+\zeta_2^2+\zeta_1\zeta_2)
\nonumber\\
&&- \frac{\Lambda}{9}(\varphi_1^2+\varphi_2^2-\varphi_1\varphi_2)-\frac{1}{3}.
\end{eqnarray}
It can be diagonalized using the generation function
$F_2=(p_1+p_2)\varphi_1-2p_2\varphi_2$, which  results in (here and in similar
formulas below we drop nonessential constant terms)
\eqb
    \tilde{\mathcal{H}}_{P_1}=\frac{1}{4}(1-3\Lambda)p_1^2 +\frac{3}{4}(1-3\Lambda)p_2^2 -
\frac{\Lambda}{12}q_1^2- \frac{\Lambda}{36}q_2^2,
\eqe
where $(p_1,q_1)$ and $(p_2,q_2)$ are the local  canonical variables. Thus $P_1$ is
unstable  for $\Lambda<\Lambda_{1}\equiv1/3$ and corresponds to saddle points on
the planes $(p_j,q_j)$, while it is linearly stable otherwise, though corresponds
to a local maximum of the Hamiltonian. The respective motion can be  interpreted as
a 2D linear oscillator with negative effective masses and with  equal frequencies
$\Omega_{P_1}=[\Lambda(3\Lambda-1)/12]^{\frac12}$.

We observe that the critical value of the lattice parameter $\Lambda_1$ for equally
populated $X$-points  coincides with that in the  2D optical lattice
(see~\cite{SK,BKKS}).

2. The second stationary point $P_2
=\{\frac13,\frac13,\frac{2\pi}{3},-\frac{2\pi}{3}\}$ also corresponds to equal
populations of the $X$-points, but characterized by mutual $2\pi/3$-phase
differences. In this case the Hamiltonian about the stationary point reads
\begin{eqnarray*}
 \mathcal{H}_{P_2} &=& \frac{1}{8}(2-3\Lambda)(\zeta_1^2+\zeta_2^2
 +\zeta_1\zeta_2)+\frac{\Lambda}{18}(\varphi_1^2+\varphi_2^2-\varphi_1\varphi_2)\\
 && +\frac{\Lambda}{4\sqrt{3}}
 [\zeta_1(2\varphi_2-\varphi_1)+\zeta_2(\varphi_2-2\varphi_1)]+\frac{\Lambda}{2}-\frac{1}{3}.
 \end{eqnarray*}
The variables $\zeta$ and $\varphi$ are now mixed and the transformation which
diagonalizes the Hamiltonian is complicated.  The eigenfrequencies, however, can be
directly obtained by considering the characteristic equation. We get two distinct
values \mbox{$\Omega_{P_2}^{(\pm)} =
\left[\frac{\Lambda}{6}\pm\sqrt{\frac{\Lambda^3}{6}-\frac{\Lambda^4}{4}}\right]^{\frac12}$}
which become complex for $\Lambda>\Lambda_2\equiv 2/3$. Therefore, the $P_2$-point
is linearly stable for $\Lambda\le\Lambda_2$ and is unstable otherwise. The
critical value  $\Lambda_2$ is a characteristic feature of the 3D case and does not
exist in the 2D setup. One also readily concludes from the symmetry that another
stationary points is given by  $P_2^\prime =
\{\frac13,\frac13,-\frac{2\pi}{3},\frac{2\pi}{3}\}$.

3. The next three stationary points are given by
$P_3=\left\{\frac{1+\Lambda}{3-\Lambda},\frac{1+\Lambda}{3-\Lambda},\pi,\pi\right\}$,
$P_4=\left\{\frac{1+\Lambda}{3-\Lambda},\frac{3\Lambda-1}{\Lambda-3},0,\pi\right\}$
and
$P_4^\prime=\left\{\frac{3\Lambda-1}{\Lambda-3},\frac{1+\Lambda}{3-\Lambda},\pi,0\right\}$.
The point $P_3$ corresponds to the populations
$$
(x_1,x_2,x_3)=\left(\frac{1-\Lambda}{3-\Lambda},\frac{1-\Lambda}{3-\Lambda},\frac{1+\Lambda}{3-\Lambda}\right).
$$
while the points $P_4$ and $P_4^\prime$ correspond to the same distributions with
$X$-points being interchanged:
$(x_1,x_2,x_3)_{P_4}=\left(\frac{1-\Lambda}{3-\Lambda},\frac{1+\Lambda}{3-\Lambda}
,\frac{1-\Lambda}{3-\Lambda}\right)$ and
$(x_1,x_2,x_3)_{P_4^\prime}=\left(\frac{1+\Lambda}{3-\Lambda},\frac{1-\Lambda}{3-\Lambda}
,\frac{1-\Lambda}{3-\Lambda}\right)$. The stability properties and the diagonalized
local Hamiltonian  are the same for these three points. Consider, for instance,
$P_3$. The local Hamiltonian
\begin{eqnarray}
 \mathcal{H}_{P_3} &=& -\frac{(1-\Lambda)^2}{3-\Lambda}+\frac14(1-\Lambda)(\zeta_1^2+\zeta_2^2)
 +\frac14(1+\Lambda)\zeta_1\zeta_2\nonumber\\
 &&+\frac{\Lambda(1-\Lambda)}{(3-\Lambda)^2}[\Lambda(\varphi_1^2+\varphi_2^2)+(1-\Lambda)\varphi_1\varphi_2]
 \end{eqnarray}
can be diagonalized by means of the canonical transformation generated by
$F_2=(p_1+p_2)\varphi_1+(p_1-p_2)\varphi_2$:
\begin{eqnarray}
     \tilde{\mathcal{H}}_{P_3}=\frac{1}{4}(3-\Lambda)p_1^2 +\frac{1}{4}(1-3\Lambda)p_2^2
     +\frac{\Lambda(1-\Lambda^2)}{4(3-\Lambda)^2}\,q_1^2
\nonumber \\
-\frac{\Lambda(1-\Lambda)(1-3\Lambda)}{4(3-\Lambda)^2}\,q_2^2.
\label{HP3}\end{eqnarray}
Hence $P_3$ is a saddle point of  the Hamiltonian in the plane $(p_2,q_2)$, and
thus is always unstable (the same is true for $P_4$ and $P_4^\prime$).

4. In the critical case, $\Lambda=\Lambda_1$, for the zero phases $\phi_{1,2}=0$
the classical Hamiltonian (\ref{EQ15}) is flat in $(z_1,z_2)$:
$\mathcal{H}(z_1,z_2,0,0)=0$ and $\dot{z}_{1,2}=0$ due to the phases, i.e. the
whole domain of $(z_1,z_2)$ has the same energy for zero phases.

\subsubsection{Boundary stationary points}

As it was mentioned above, there exist boundary stationary points corresponding to
all atoms populating only one X-point  or only two X-points. As  the phases become
undefined in such a case, it is convenient to use the semi-classical Hamiltonian
obtained directly from Hamiltonian (\ref{Heff}) by the substitution
$b_j^\dag\to\sqrt{N}{b^*}^{(cl)}_j$ and $b_j\to \sqrt{N}b^{(cl)}_j$ (to have
normalized amplitudes).

1. Consider the solutions with $b^{(cl)}_{1,2} =\beta_{1,2}$ and $b_3 = 1
+\beta_3$, $|\beta_j|\ll1$, i.e. close to the stationary point $P_{B_1}$ describing
all atoms occupying just one X-point (X$_3$-point in this case):
\eqb
(x_1,x_2,x_3)=(0,0,1).
\eqe
Using that $|1+\beta_3|^2 = 1 - |\beta_1|^2 -|\beta_2|^2$ we obtain the local
Hamiltonian as follows
\begin{eqnarray}
\mathcal{H}_{X_3} &=& \mathcal{H}^{(1)}_{X_3} +\mathcal{H}^{(2)}_{X_3},\nonumber\\
\mathcal{H}^{(j)}_{X_3}& =& \frac18
+\left(\Lambda-\frac{1}{2}\right)\beta_j^*\beta_j +\frac{\Lambda}{4}[(\beta_j^*)^2
+\beta_j^2].\label{HamX3}
\end{eqnarray}
The eigenfrequencies are equal for the two modes $\beta_{1,2}$: $\Omega_{B_1} =
\frac{1}{2}[3\Lambda^2-4\Lambda +1]^{\frac12}$.  Hence for $\Lambda\le\Lambda_1$
this stationary point is a local minimum and is stable, while for
$\Lambda>\Lambda_1$ it is  unstable.

2. Moreover, it is easy to show that for $\Lambda\ge \Lambda_1$ there is one more
stationary point $P_{B_2}$ with  two X-points being equally populated (see
Appendix B). It reads:
\eqb
(x_1,x_2,x_3)=\left(\frac12,\frac12,0\right),\quad  \phi_1 = \phi_2 = \phi_\Lambda,
\label{SB2}\eqe
where $\cos(\phi_\Lambda) = \frac{1-\Lambda}{2\Lambda}$. In terms of $z$ we have
\mbox{$P_{B_2} = \{0,0,\phi_\Lambda,\phi_\Lambda\}$.} Noticing that this point is
also a stationary point of the Hamiltonian (\ref{EQ15}) we obtain the local
Hamiltonian as follows
\begin{eqnarray}
\mathcal{H}_{P_{B_2}} &=& \frac{1}{8}(1-3\Lambda)(2 +
\zeta_1^2+\zeta_2^2)-\frac{\Lambda}{8}(\varphi_1-\varphi_2)^2\nonumber\\
&-&\frac{\Lambda}{4}\sin\phi_\Lambda(\zeta_1+\zeta_2)(\varphi_1+\varphi_2)
\label{HPB2}\end{eqnarray}
Next, using the local canonical transformation with the generating function  $F_2 =
(p_1+p_2)\varphi_1 +(p_1-p_2)\varphi_2$  we arrive at (dropping the constant)
\begin{eqnarray}
&&\mathcal{H}_{P_{B_2}} = \frac{1}{4}(1-3\Lambda)(p_1^2+p_2^2)
-\frac{\Lambda}{2}\sin\phi_\Lambda p_1q_1 - \frac{\Lambda}{8}q_2^2\nonumber\\
&& = \frac{1}{4}(1-3\Lambda)\left[\tilde{p}_1^2 +p_2^2 -
\frac{\Lambda^2\sin^2\phi_\Lambda}{(1-3\Lambda)^2}q_1^2\right]-
\frac{\Lambda}{8}q_2^2,
\end{eqnarray}
with $\tilde{p}_1 = p_1-\frac{\Lambda\sin\phi_\Lambda}{1-3\Lambda}q_1$. Therefore,
$P_{B_2}$ is a saddle point in the whole domain of its
existence (except in the critical case $\Lambda=1/3$), hence it is unstable unless
$\Lambda = \Lambda_1$.

For the sake of convenience, in Table~\ref{tab:1} we present the list of the
stationary point and their linear stability properties.
\begin{table}[htbp]
    \centering
        \begin{tabular}{|l|l|l|l|}
        \hline
  &  \begin{minipage}{0.8cm} coordinates
        \newline
        $ \{z_1,z_2,\phi_1,\phi_2 \}$
        \end{minipage}
        & stability &  \begin{minipage}{0.7cm}
    populations
    \newline
          $(x_1,x_2,x_3)$
          \end{minipage}
          \\
          \hline
        $P_1$ & $\left\{ \frac 13,\frac 13,0,0\right\}$  & stable for $\Lambda>\frac 13$    &   $\left(\frac 13,\frac 13,\frac 13\right)$
        \\
        $P_2$ & $\{\frac13,\frac13,\frac{2\pi}{3},-\frac{2\pi}{3}\}$  & stable for $\Lambda<\frac 23$   &   $\left(\frac 13,\frac 13,\frac 13\right)$
        \\
        $P_2^\prime$ & $\{\frac13,\frac13,-\frac{2\pi}{3},\frac{2\pi}{3}\}$  & stable for $\Lambda<\frac 23$    &   $\left(\frac 13,\frac 13,\frac 13\right)$
        \\
        $P_3$ & $\left\{ \frac{1+\Lambda}{3-\Lambda},\frac{1+\Lambda}{3-\Lambda},\pi,\pi\right\}$  & unstable   &   $\left(\frac{1-\Lambda}{3-\Lambda},\frac{1-\Lambda}{3-\Lambda},\frac{1+\Lambda}{3-\Lambda}\right)$
        \\
        $P_4$ & $\left\{ \frac{1+\Lambda}{3-\Lambda},\frac{3\Lambda-1}{\Lambda-3},0,\pi\right\}$  & unstable    &   $\left(\frac{1-\Lambda}{3-\Lambda},\frac{1+\Lambda}{3-\Lambda}
,\frac{1-\Lambda}{3-\Lambda}\right)$
        \\
        $P_4^\prime$ & $\left\{ \frac{3\Lambda-1}{\Lambda-3},\frac{1+\Lambda}{3-\Lambda},\pi,0\right\}$  & unstable     &   $\left(\frac{1+\Lambda}{3-\Lambda},\frac{1-\Lambda}{3-\Lambda}
,\frac{1-\Lambda}{3-\Lambda}\right)$
        \\
        $P_{B_1}$ & not used  & stable for $\Lambda<\frac 13$   &   $(0,0,1)$
        \\
        $P_{B_2}$ & not used  & unstable for $\Lambda\neq \frac 13$     &   $\left(\frac 12,\frac 12,0\right)$
        \\
        \hline
        \end{tabular}
    \caption{The stationary points of the classical Hamiltonian and their linear stability properties}
    \label{tab:1}
\end{table}


\section{Quantum evolution}
\label{numerics}

\subsection{The initial state and the numerical approach}

We are interested in quantum dynamics of an initial state of a large number of
atoms with the average values of the occupation numbers and the phases being close
to a semi-classical stationary point. To find out the structure of such an initial
state consider the semi-classical wave function $\psi(x_1,x_2,\tau) =
e^{iS(x_1,x_2,\tau,h)/h}$, where in the lowest-order approximation:
 $S(x_1,x_2,\tau,h) = S^{(cl)}(x_1,x_2,\tau) +\mathcal{O}(h)$. In the vicinity of
a stationary point $(x^{(cl)}_1,x^{(cl)}_2)$ we can expand the classical action as
follows
\begin{eqnarray}
S^{(cl)} &=& -E^{(cl)}\tau + \phi^{(cl)}_1(x_1-x^{(cl)}_1)
+\phi^{(cl)}_2(x_2-x^{(cl)}_2)
\nonumber\\
&& + \mathcal{O}[(x_1-x^{(cl)}_1)^2 + (x_2-x^{(cl)}_2)^2].
\label{ACTEXP}\end{eqnarray}
Therefore, recalling that $x_j = k_j/N$ and $h = 2/N$ and taking into account that
the average values of $x_{1,2}$ must be close to the semi-classical ones
$x^{(cl)}_{1,2}$, we can approximate the initial state by the Gaussian function
\eqb
C_{k_1,k_2} = C_0e^{\frac{i}{2}(\phi^{(cl)}_1k_1+\phi^{(cl)}_2k_2) -
\frac{\left(k_1-k^{(cl)}_1\right)^2 +\left(k_2-k^{(cl)}_2\right)^2}{2\sigma_N^2}}.
\label{INCOND}\eqe
Here $\phi^{(cl)}_{1,2}$ and $k^{(cl)}_{1,2}$ are the classical phases  and
populations, $C_0$ is the normalization factor and $\sigma_N$ is the width
parameter such that
\eqb
1 \ll\sigma_N\ll N
\label{CONDs}
\eqe
(the first inequality is imposed to guarantee smoothness of $S(x_1,x_2,\tau,h)$
with respect to $x_{1,2}$ and the second one is the condition of small width of the
wave-packet in the Fock space). Due to symmetry of the quantum Hamiltonian (bosons
are created by pairs), the classical phases $\phi_{1,2}$ give rise to six different
quantum states of the form (\ref{INCOND}) with the phases $\phi_{1,2}+ 2\pi
s_{1,2}$, $s_{1,2}\in \{-1,0,1\}$ (see also the discussion of phase states below).

One can expect that the state (\ref{INCOND}), (\ref{CONDs}) with the  classical
variables satisfying the respective Hamiltonian equations is a good approximation
for the actual quantum state for all times $\tau$ as $h\to0$  if the classical
stationary point is stable. Indeed,  in this case the expansion (\ref{ACTEXP}) can
be truncated as indicated.

To get a  numerical solution of Schr\"odinger equation (\ref{EQ10}) with a
controllable accuracy we have used the method of Ref. \cite{TalEzerKosloff}, i.e.
the expansion of the unitary operator $U = \exp\{-iN\hat{H}\tau\}$ over the
Chebyshev polynomials
\eqb
e^{-iN\hat{H}\Delta \tau} = e^{-iN\bar{E}\tau}\sum_{\ell=0}^\infty C_\ell(N\Delta
E\Delta \tau )T_\ell(\hat{I}),
\label{CHEB}\eqe
where $\bar{E} = (E_\mathrm{max} + E_\mathrm{min})/2$, $\Delta E = (E_\mathrm{max}
- E_\mathrm{min})/2$, with  $E_\mathrm{min}$ and $E_\mathrm{max}$ being the lower
and upper bounds taken from equation (\ref{EST}), $T_\ell(\hat{I})$ being the
$\ell$-order Chebyshev polynomial of the Hermitian operator \mbox{$\hat{I} =
(\hat{H} - \bar{E})/\Delta E$} with the eigenvalues lying on the interval $[-1,1]$.
The coefficients are given as $C_\ell(\varkappa) =
(-i)^\ell(2-\delta_{\ell,0})J_\ell(\varkappa)$ where $J_\ell(\varkappa)$ is the
Bessel function of the first kind.    Due to the uniform convergence  of the
Chebyshev series on $[-1,1]$ and the fact that the coefficients vanish
exponentially for sufficiently large $\ell$ (for a fixed $\Delta \tau$) one can
compute the evolution operator for the Schr\"odinger equation at the  times $\tau =
\Delta \tau, 2\Delta \tau, 3\Delta \tau, \ldots$ with arbitrary given accuracy,
limited only by the roundoff errors (we have set the error to be of the order
$10^{-8}$).

\subsection{Quantum evolution about the $P_1$-state}

\begin{figure}[htb]
\begin{center}
\vskip 1cm
\includegraphics[scale=0.5, bb = 65   245   445   619 ]{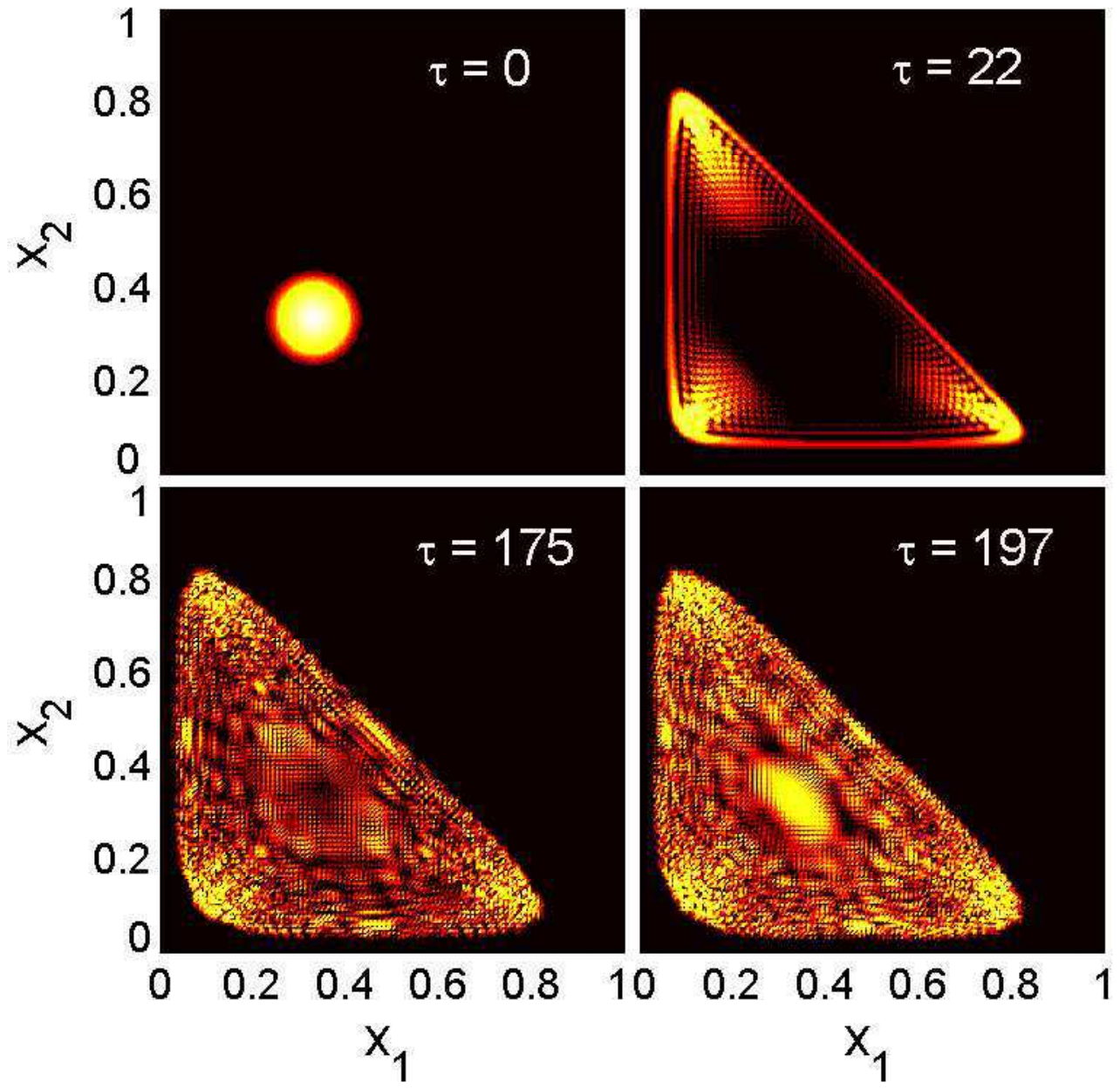}
\caption{Quantum evolution of $N=200$ BEC atoms loaded into the high symmetry
points $X_1$, $X_2$ and $X_3$. The lattice constant is $\Lambda=0.21$. We use the
initial state as in Eq.(\ref{INCOND}) with $\sigma = \sqrt{N/2}$ with the initial
populations $(x_1,x_2)=(0.330,0.337)$ and phases $(\phi_1,\phi_2)=(0.02,-0.02)$.
The initial stage of evolution is given in the upper two panels, while the lower
two panels show an oscillating  state by which the initial (localized) state is
replaced. }
\label{FG1}
\end{center}
\end{figure}

For $\Lambda<\Lambda_1$ $P_1$ is a saddle point and is unstable with respect to
small perturbations. An initial quantum state in the form (\ref{INCOND}) such that
the average initial populations $x_j$ and the phases $\phi_j$ are close to the
semi-classical stationary values $x_j=1/3$ and $\phi_j=0$ results in the evolution
presented in Fig. \ref{FG1}. The initial localized, nearly-Fock, state  transforms
to a broad oscillating state (lower panels of Fig. \ref{FG1}) persisting at least
for some long evolution time.

The emergent state can be approximated by a linear combination of a small number of
the phase states. The one-dimensional phase states are defined here via the
discrete Fourier transform (DFT)
\eqb
|\theta_\ell\rangle = \frac{1}{\sqrt{N+1}}\sum_{n=0}^Ne^{in\theta_\ell}|n\rangle,
\label{PhST}\eqe
where $\theta_\ell = \frac{2\pi\ell}{N+1}$. Evidently $\langle
\theta_{\ell^\prime}|\theta_\ell\rangle = \delta_{\ell^\prime,\ell}$. Therefore,
the phase states give another basis of the Hilbert space, in fact
\eqb
|n\rangle =
\frac{1}{\sqrt{N+1}}\sum_{\ell=0}^Ne^{-in\theta_\ell}|\theta_\ell\rangle.
\label{Fock}\eqe

For a fixed  total number of atoms $n_1+n_2+n_3=N$ the three-dimensional phase
states are projected onto a two-dimensional subspace, i.e. the wave function can be
written as
\eqb
|\Psi\rangle =
\sum_{\ell_1=0}^N\sum_{\ell_2=0}^N\hat{C}_{\ell_1,\ell_2}|\theta_{\ell_1},\theta_{\ell_2}\rangle,
\label{Phases}\eqe
where
\eqb
\hat{C}_{\ell_1,\ell_2} \equiv
\frac{1}{N+1}\sum_{k_1=0}^N\sum_{k_2=0}^{N-k_1}e^{-ik_1\theta_{\ell_1}-ik_2\theta_{\ell_2}}C_{k_1,k_2}
\label{FFTC}\eqe
is nothing but the DFT of the coefficients
$C_{k_1,k_2}$ extended over whole domain of $0\le k_{1,2}\le N $ by padding them
with zeros. Note that the phase  $\theta$ is half of the value of the
semi-classical phase $\phi$ in the limit $h\to0$.

\begin{figure}[htb]
\begin{center}
\includegraphics[scale=0.5, bb = 48   296   478   555]{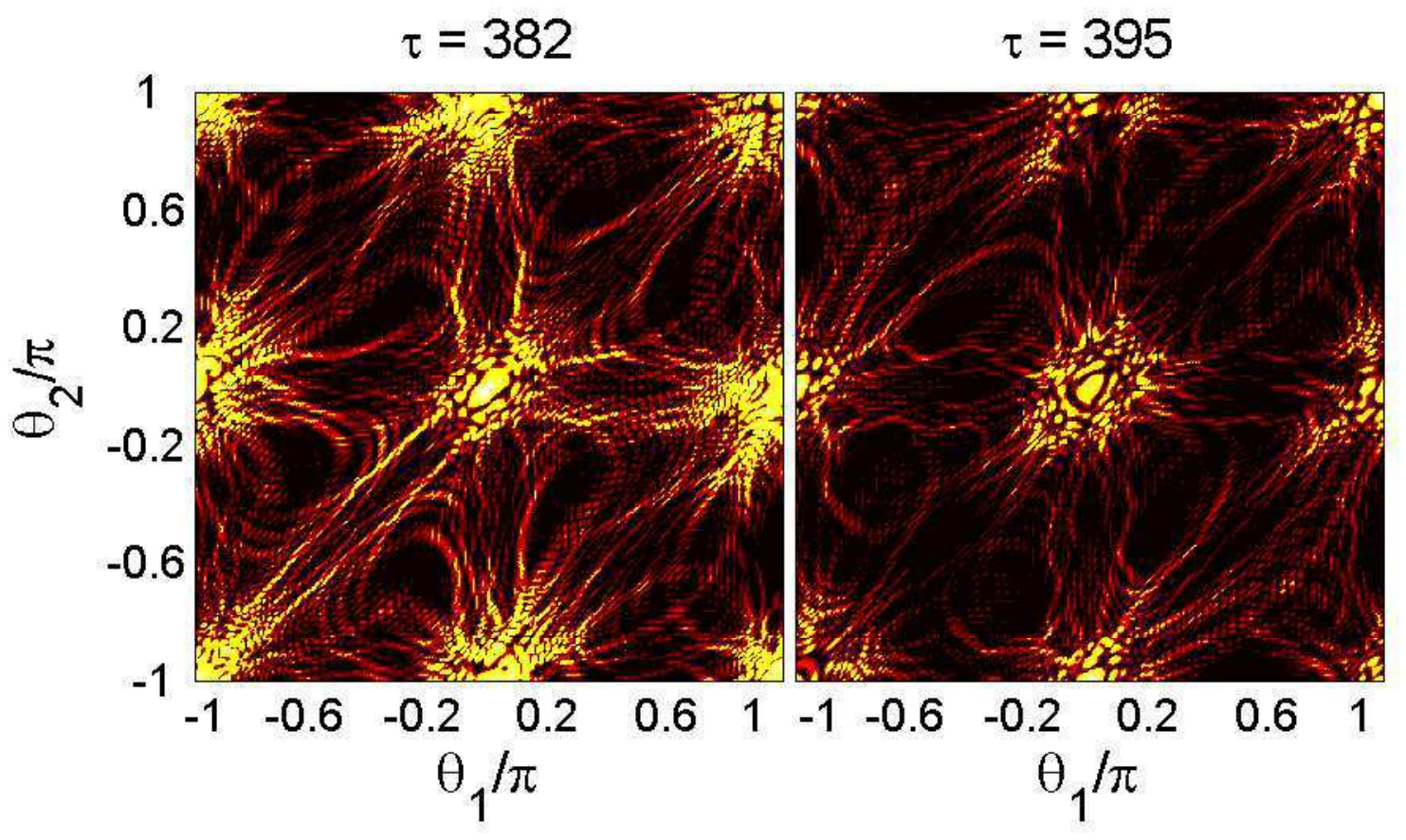}
 \caption{The DFT
transform of the wave function of Fig. \ref{FG1} at two large times (two panels are
used to show the relatively small deformation with time). }
\label{FG2}
\end{center}
\end{figure}

We find that the DFT of the wave function of Fig. \ref{FG1} is concentrated at the
following values of the phases $\theta_\ell = \{0,\pm\pi\}$, see Fig. \ref{FG2}.
These states correspond to the phases $\phi_j=0$ in the semi-classical limit, i.e.
to the phases of the stationary point $P_1$.

\begin{figure}[htb]
\begin{center}
\includegraphics[scale=0.5, bb = 75   244   443   649 ]{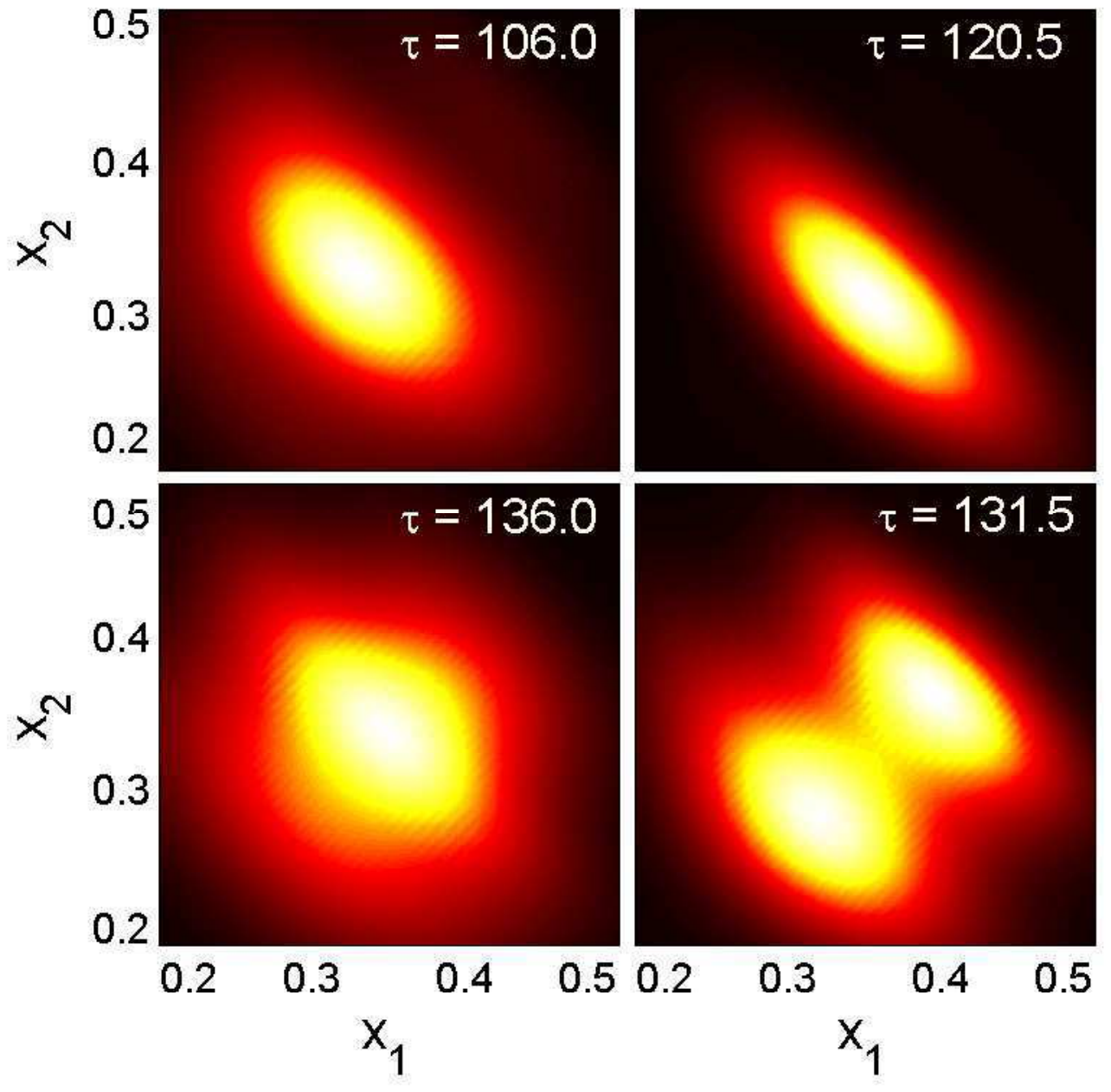} \caption{Quantum evolution of an initial Gaussian state of
$N=200$ BEC atoms with $\sigma = \sqrt{N}$, the initial populations
$(x_1,x_2)=(0.35, 0.32)$, and phases $(\phi_1,\phi_2)=(0.01,-0.02)$. The lattice
constant $\Lambda=0.36$, i.e. the semi-classical state $P_1$ is stable.
Oscillations of the  wave function about the initial state are observed (the time
increases clock-wise).   }
\label{FG3}
\end{center}
\end{figure}

\begin{figure}[htb]
\begin{center}
\includegraphics[scale=0.5, bb = 85   253   410   585]{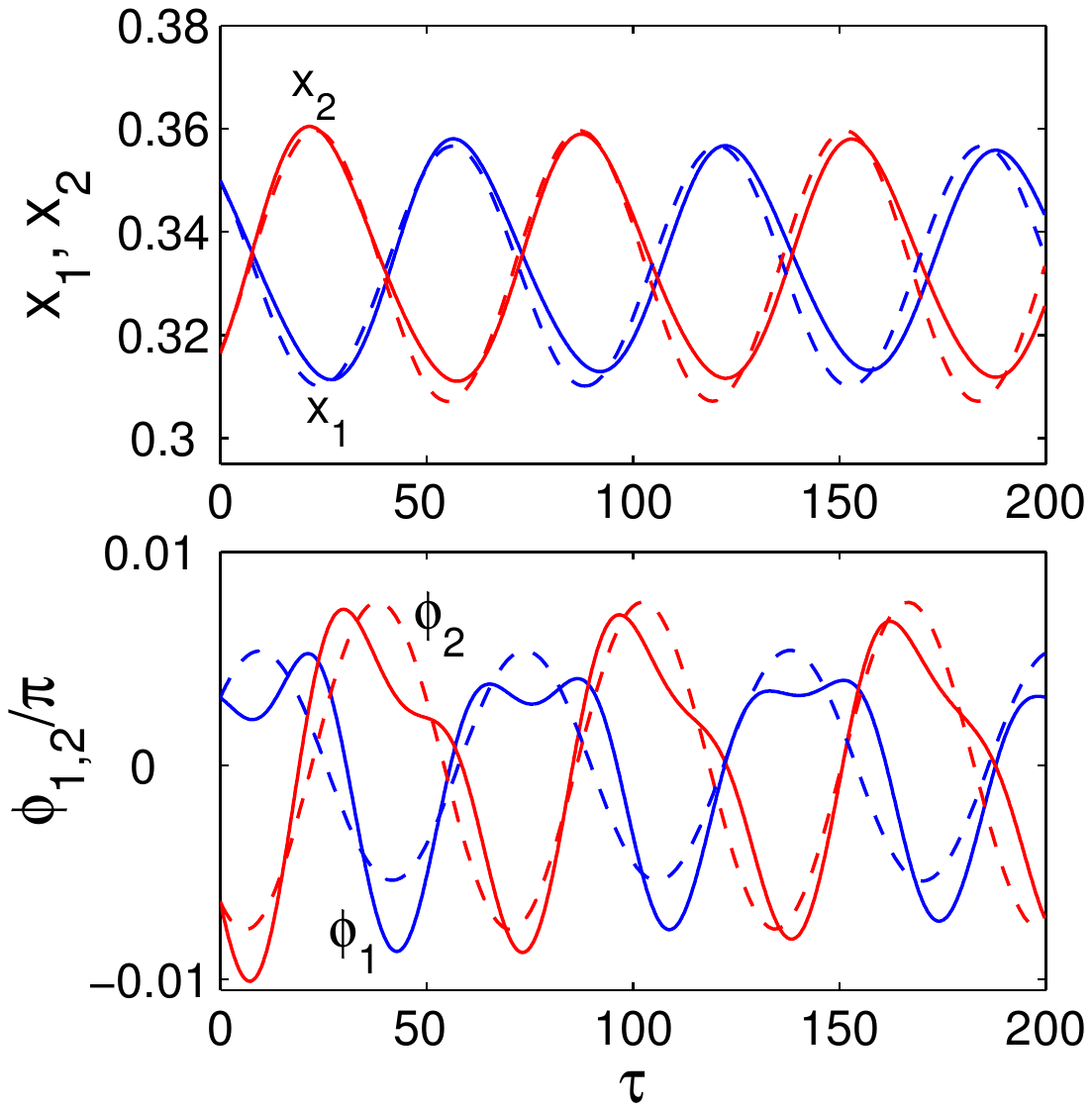}
\caption{Comparison of the quantum evolution of Fig. \ref{FG3} with the
semi-classical evolution corresponding to the initial average values of the
populations and phases. The upper and lower panels show average populations and phases, respectively.}
\label{FG4}
\end{center}
\end{figure}

The quantum evolution of the initial state corresponding to a stable classical
stationary point as $h\to0$ is different, see Fig. \ref{FG3}. First of all, the
localized (i.e. nearly Fock) state remains localized. Note that the quantum and the
semi-classical dynamics are very close in this case, see Fig. \ref{FG4}, though the
number of atoms is rather small.

In the nonlinear tunneling of BEC in a square 2D optical lattice \cite{SK}  (where
the two-mode model appears) the quantum evolution features appear as collapses and
revivals of the semi-classical dynamics. The  energy spacing $\delta E \sim N^{-2}$
discussed in in Sec. \ref{model} for the three-mode model (as compared to $\delta E
\sim N^{-1}$ for the two-mode model) prevents observation of the quantum collapse.
Indeed, the semi-classical regime requires large number of atoms, thus large
evolution times $\tau \sim N^2$ are required for observation of the first quantum
collapse. We verified that the quantum oscillations of Fig. \ref{FG4} follow the
semi-classical ones without occurrence of the quantum collapse for times up to
$\tau = 60 000$ at least, which would exceed by far the lifetime of BEC (see
Sec.~\ref{model}). This result also suggests that the quantum collapse may not
exist in the model at all.

\subsection{Quantum evolution about the $P_2$-state}

The above results show that the quantum model of $N$ identical bosons distinguishes
between the stable and unstable classical stationary points. This conclusion agrees
with the correspondence between the quantum stability of a semi-classical state in
a system of identical bosons and the Hamiltonian stability of the corresponding
stationary point in the classical limit \cite{QStab}.

\begin{figure}[htb]
\begin{center}
\includegraphics[scale=0.5, bb = 65   245   442   649]{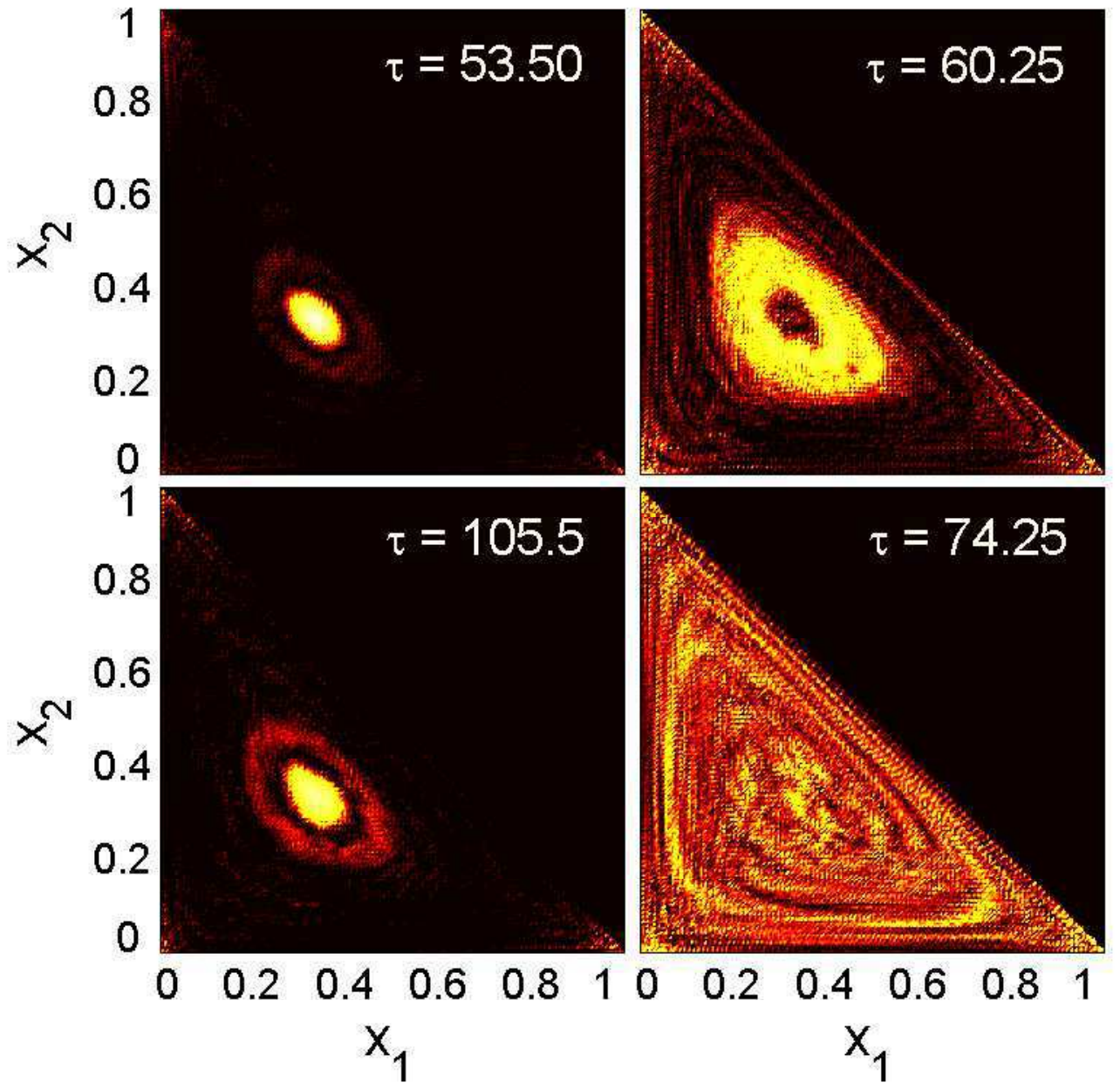} \caption{Recurrence of the wave function in quantum evolution
(the time increases clock-wise) of an initial Gaussian state of $N=200$ BEC atoms
with $\sigma = \sqrt{N}$, the initial populations $(x_1,x_2)=(0.330, 0.337)$, and
phases $(\phi_1,\phi_2)=\pi(2/3+0.01, -2/3+0.006)$. The lattice constant
$\Lambda=0.64$, i.e. the semi-classical state $P_2$ is stable.   }
\label{FG5}
\end{center}
\end{figure}

\begin{figure}[htb]
\begin{center}
\includegraphics[scale=0.5, bb = 79   253   410   585]{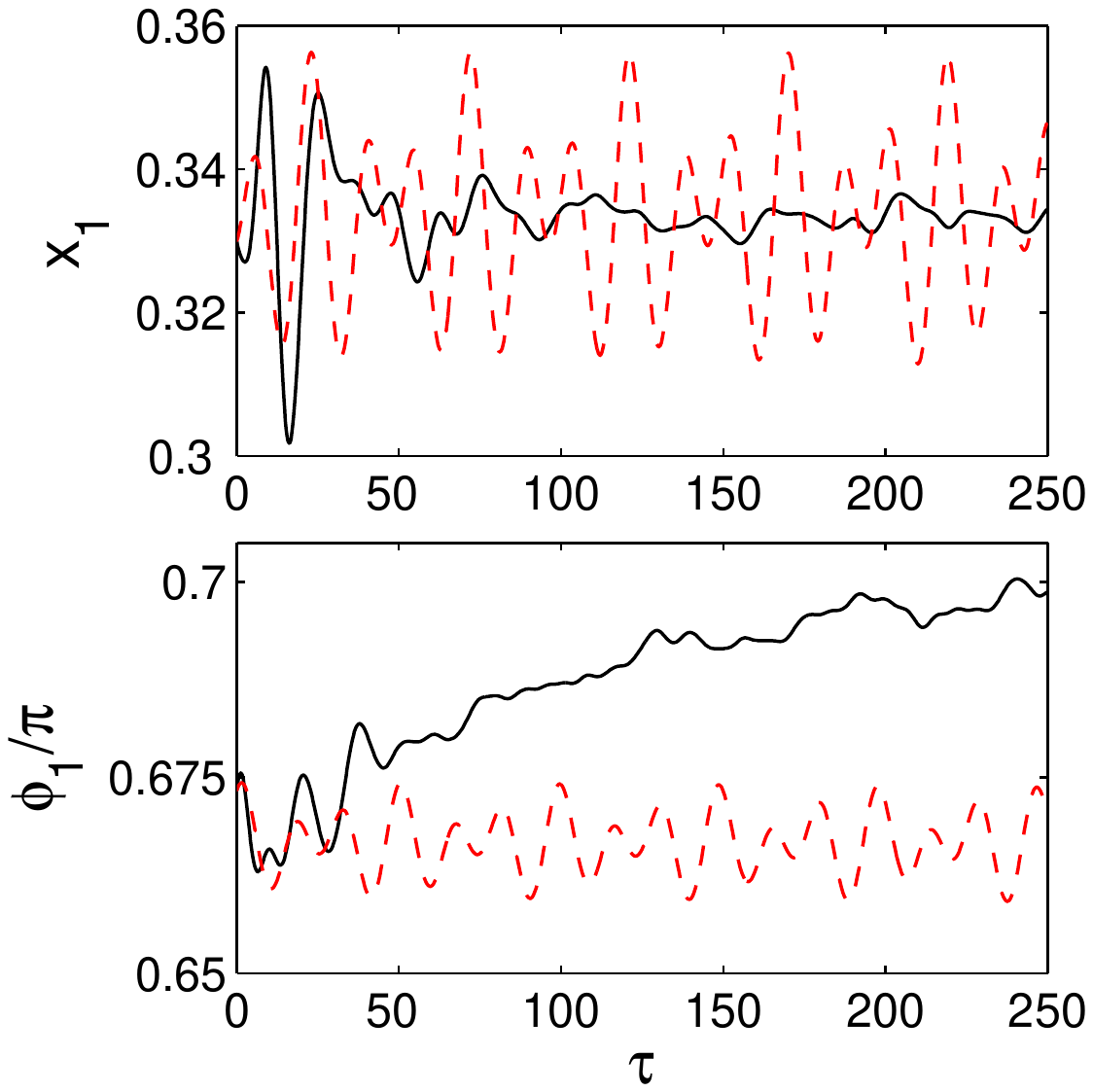}
\caption{Comparison of the quantum evolution  of Fig. \ref{FG5} (solid lines) with
the  semi-classical result corresponding to the initial average values of the
populations and phases. The upper panel gives the average populations and the lower
one the phases (dashed lines). Top panel gives the population $x_1$ and the bottom
one the phase $\phi_1$. }
\label{FG6}
\end{center}
\end{figure}

However, due the discreteness of the quantum energy levels  the quantum evolution
can have features not found in the classical model (the two cases, of course, agree
in the limit $N\to \infty$ when the quantum energy spacing goes to zero). This is
clearly illustrated by the results presented in Figs. \ref{FG5} and \ref{FG6}.
Indeed, Fig. \ref{FG5} illustrates one period of the wave-function spread and
subsequent recurrence to the localized distribution, which is responsible for the
deviation of the quantum averages from the corresponding classical variables, see
Fig. \ref{FG6}. Note however, that the quantum averages remain close to the the
classical stationary point values, in accordance with the general correspondence of
the quantum and classical stability  \cite{QStab}.

One more difference is apparent in Fig. \ref{FG6} as compared with Fig. \ref{FG4}:
the semi-classical dynamics about the $P_2$-point features two frequencies instead
of one, as it is for the $P_1$-point. Despite the disagreement of the quantum
averages and the classical dynamics, the recurrence period is in fact very close to
one of the classical oscillations periods $\tau \sim 50$.

\begin{figure}[htb]
\begin{center}
\vskip 2cm
\includegraphics[scale=0.5, bb = 62   296   490   542]{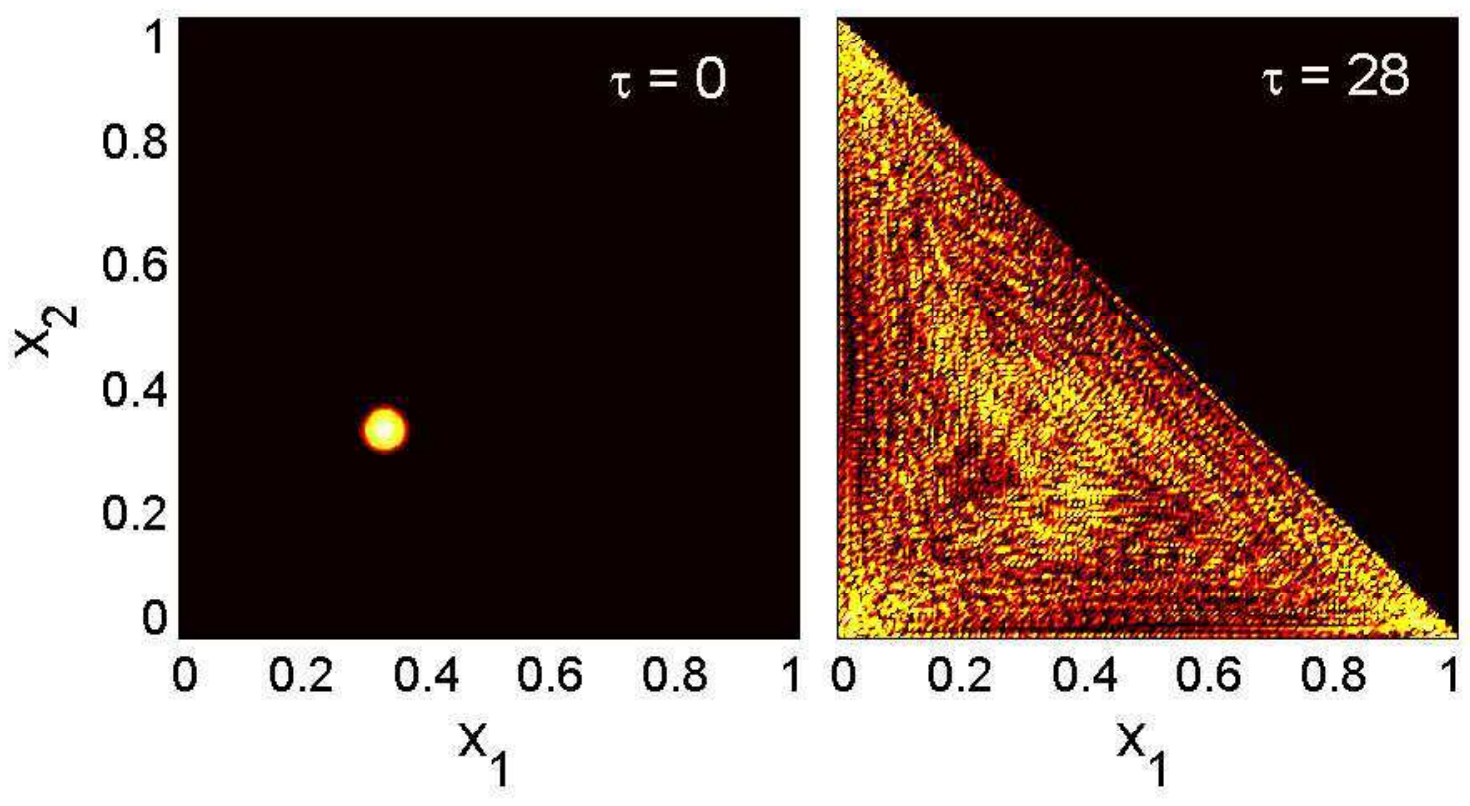} \caption{Quantum
evolution of an initial Gaussian state of $N=200$ BEC atoms with $\sigma =
\sqrt{15}$, the initial populations $(x_1,x_2)=(0.330 0.337)$, and phases
$(\phi_1,\phi_2)=\pi(2/3+0.07, -2/3+0.07)$. The lattice constant $\Lambda=0.69$ and
the semi-classical state $P_2$ is unstable. }
\label{FG7}
\end{center}
\end{figure}

\begin{figure}[htb]
\begin{center}
\includegraphics[scale=0.5, bb = 48   296   490   555]{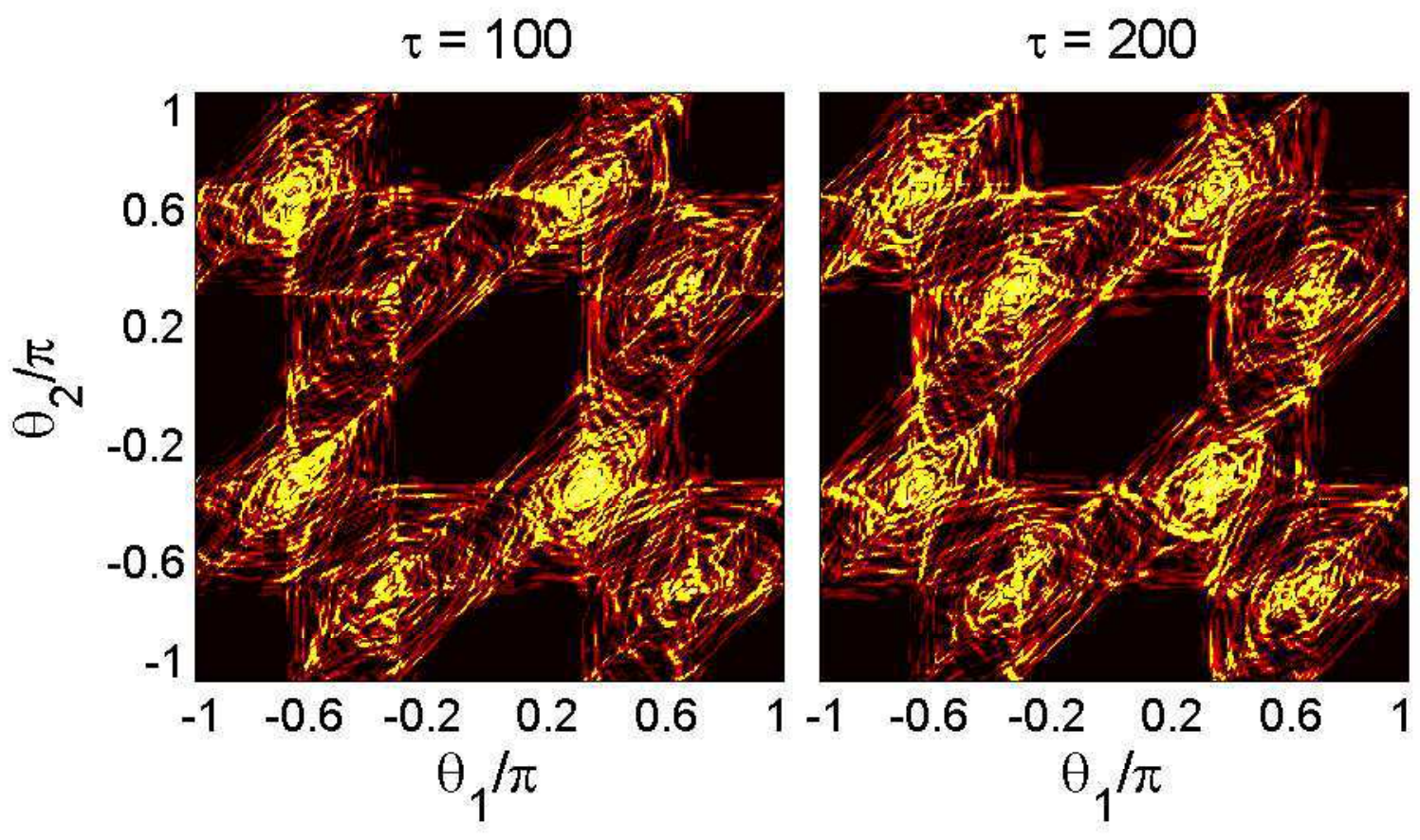} \caption{The DFT
transform of the wave function represented in the right panel of Fig. \ref{FG7}. }
\label{FG8}
\end{center}
\end{figure}

The  quantum dynamics corresponding to the unstable classical fixed point $P_2$ is
similar to that in the case of unstable $P_1$-point, namely the localized, i.e.
nearly Fock-state, is replaced by a linear combination of a small fraction of the
phase states, see Figs. \ref{FG7} and \ref{FG8}. The phase states of Fig. \ref{FG8}
are concentrated about the following phases:
\begin{eqnarray*}
&& \biggl\{\left(-\frac{2\pi}{3},-\frac{\pi}{3}\right),
\left(-\frac{2\pi}{3},\frac{2\pi}{3}\right),
\left(-\frac{\pi}{3},\frac{\pi}{3}\right),
\left(\frac{\pi}{3},\frac{2\pi}{3}\right),
\\
&& \left(-\frac{\pi}{3},-\frac{2\pi}{3}\right),
\left(\frac{2\pi}{3},-\frac{2\pi}{3}\right),
\left(\frac{\pi}{3},-\frac{\pi}{3}\right),
\left(\frac{2\pi}{3},\frac{\pi}{3}\right)\biggr\},
\end{eqnarray*}
which correspond to the classical phases
$(\phi_1,\phi_2)=\left\{\left(\pm\frac{2\pi}{3},\mp\frac{2\pi}{3}\right)\right\}$,
i.e. to the phases of the stationary point $P_2$ and its equivalent $P^\prime_2$.

For a special initial atomic distributions it is possible to have stable-like
quantum dynamics about an unstable semi-classical fixed point which is
conditionally stable for the special initial conditions. For instance, the fixed
point $P_3$ is stable for the initial states with no $(p_2,q_2)$-components in the
classical limit (which is supposed to be a small perturbation about the fixed
point). In this case the wave function remains localized and performs oscillations
about the initial state (not shown).

As the stationary points corresponding to unequal populations of the $X$-points of
the lattice are unstable and loading BEC into the unequal distribution among the
high-symmetry points is not an easy (if at all possible) task we discard the
further  analysis of the dynamics about the points $P_3$, $P_4$ and $P^\prime_4$.

\subsection{Dynamics of the boundary states}

\begin{figure}[htb]
\begin{center}
\includegraphics[scale=0.5, bb = 38   253   410   585]{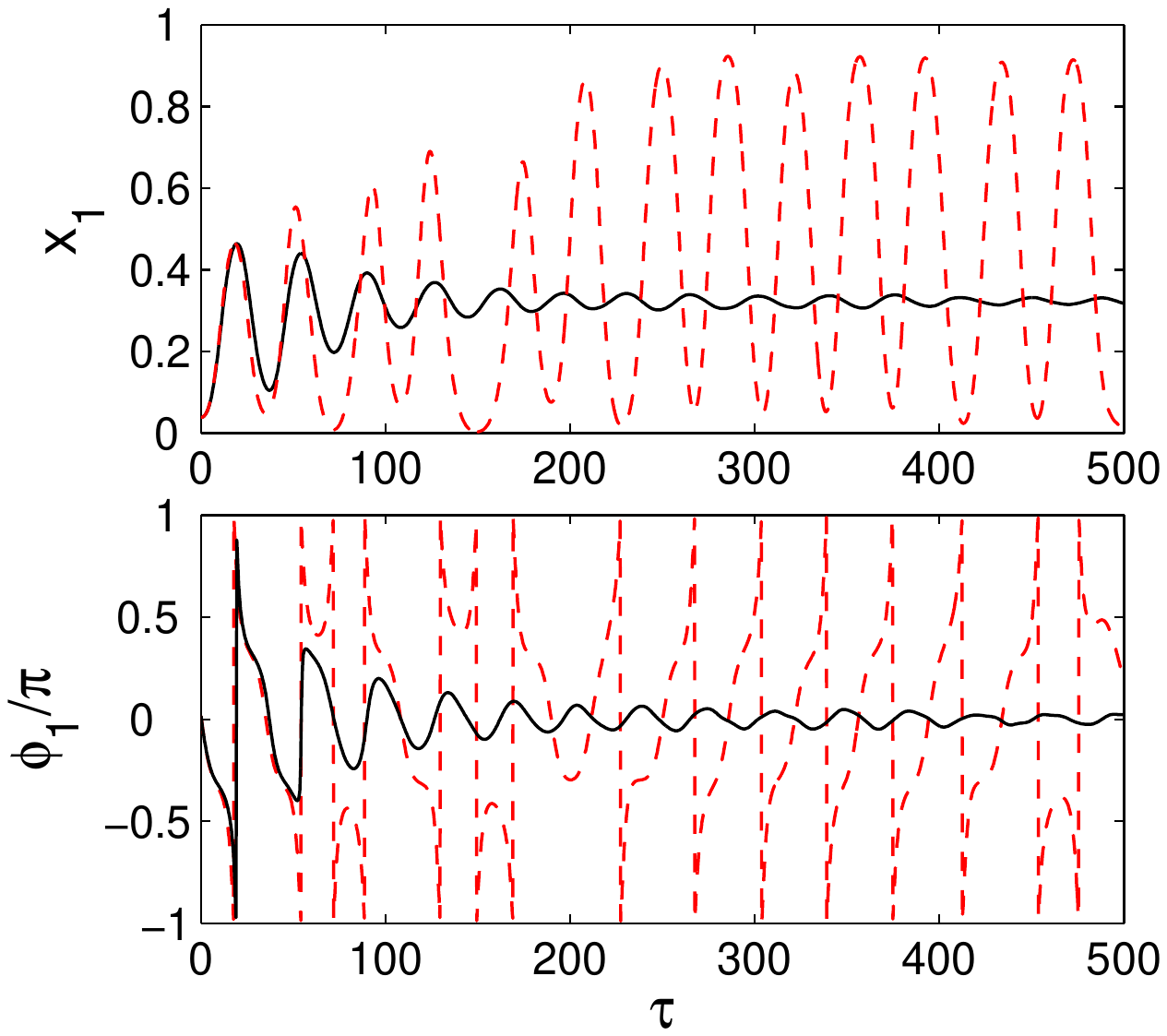}
\caption{Comparison of the quantum evolution (solid lines) of  with the
semi-classical one (dashed lines). The upper panel gives the average population
$x_1$ and the lower one the phase $\phi_1$.  Here the initial populations and
phases are $(x_1,x_2) = (0.04,0.04)$ and $(\phi_1,\phi_2) = (0.05,0)$, the lattice
parameter $\Lambda = 0.41$, $N = 200$, and $\sigma = \sqrt{N}$. The stationary
point $P_{B_1}$ ($x_3=1$) is unstable. }
\label{FG9}
\end{center}
\end{figure}

One can easily load BEC into a single $X$-point by switching on a moving cubic
lattice. Thus, it is important to consider the boundary stationary point $P_{B_1}$.
Let us consider X${}_3$-point being initially populated. For $\Lambda<1/3$ the
point  $P_{B_1}$ is classically stable and the quantum dynamics consists of
localized oscillations about the initial state. If however, the lattice parameter
passes the critical value $\Lambda_1$ the instability of $P_{B_1}$ results in
tunneling to the equal distribution of atoms between the three X-points, see Fig.
\ref{FG9}.

The dynamical instability of the $P_{B_1}$ can be used to prepare the system in the
equal distribution of atoms between the $X$-points by loading first the
$P_{B_1}$-point as discussed above and modifying the lattice parameter $\Lambda$ to
force the dynamical instability of $P_{B_1}$ to develop, i.e. as shown in Fig.
\ref{FG9}. The oscillations in Fig. \ref{FG9} are about an equal distribution of
atoms between the $X$-points and the zero phases, thus the quantum state is the
semiclassical state about the $P_1$-point of the form given by equation
(\ref{INCOND}) (or a linear combination of such states). Moreover, one can notice
that the energy of a stationary semiclassical state ($N\gg1$) in the main order is
given by the zero-point energy of the local classical Hamiltonian, since the energy
spacing between the local bound states is on the order or smaller than $1/N$ (since
the quantum oscillator model, obtained by the ``reverse quantization'' procedure of
the local classical Hamiltonian, has the energy spacing $O(h)$). Comparing the
energies $E_{P_2,P_2^\prime} = \Lambda/2 - 1/3 +O(1/N)$ and $E_{P_1} = -1/3
+O(1/N)$, we see that the ground state for $\Lambda \ge 1/3$ corresponds to the
$P_1$-point.

On the other hand, the $P_2$-point and its equivalent point $P_2^\prime$ are stable
and $P_1$ is unstable for $\Lambda < 1/3$ (see table \ref{tab:1}). Note also that
the zero-point energy of $E_{P_2} = \Lambda/2 < 1/6$ is lower than that of
$E_{P_{B_1}} = 1/4$, another stable point for $\Lambda <1/3$.  Thus, given the
quantum state with an equal distribution between the $X$-points, i.e. $P_1$, one
can prepare another such stable state (in fact $P_2$ or $P_2^\prime$ or their
linear combination) by repeating the above procedure but now starting from the
$P_1$-point by adiabatically changing the lattice parameter to $\Lambda<1/3$
followed by the thermal cooling procedure.

\begin{figure}[htb]
\begin{center}
\includegraphics[scale=0.5, bb = 70   245   442   649]{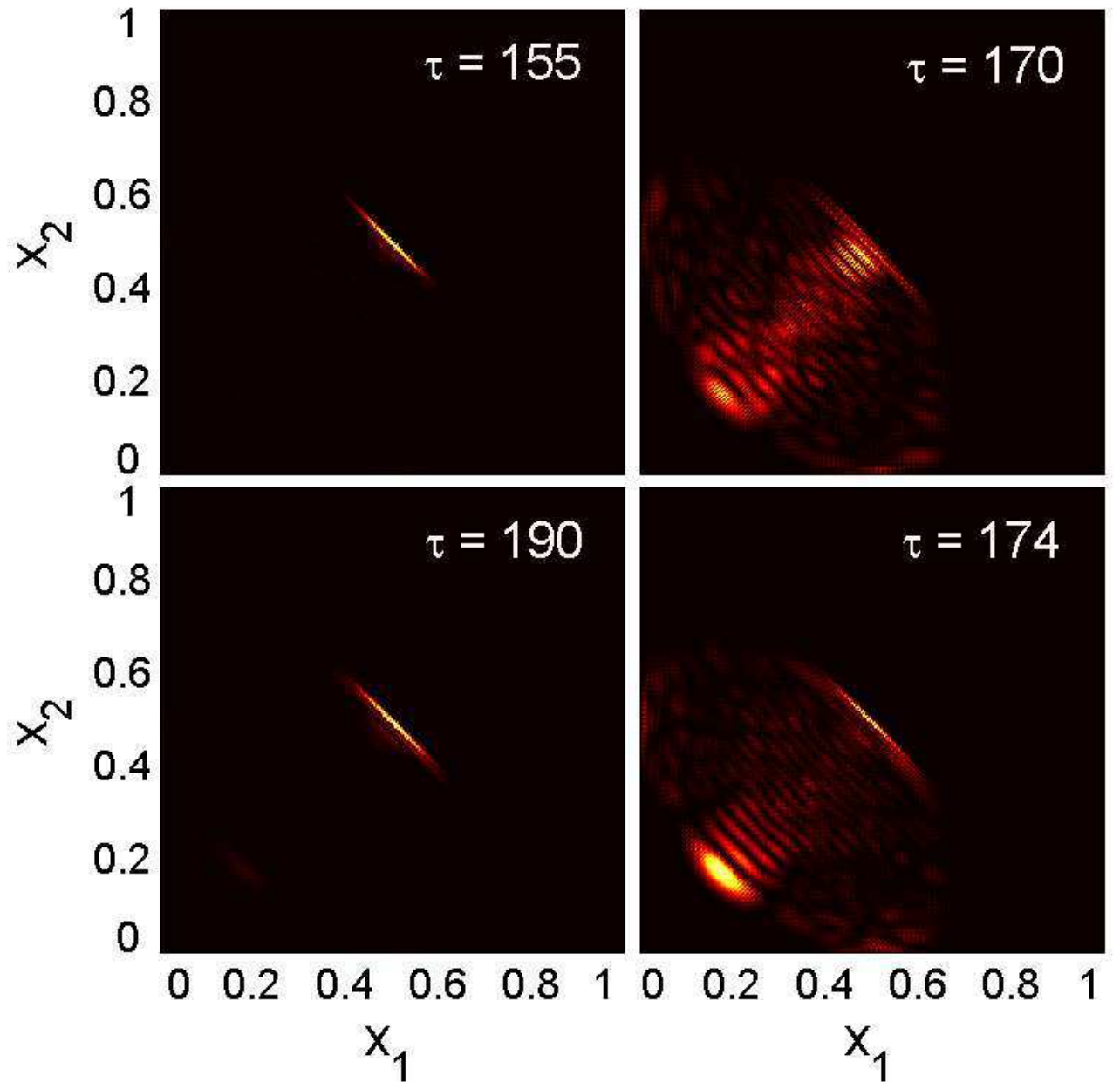} \caption{Quantum
evolution of a state corresponding to initially almost equally populated two
X-points (here X${}_1$ and X${}_2$). The time increases clock-wise. Here the
initial populations and phases are $(x_1,x_2) = (0.48, 0.48)$ and $(\phi_1,\phi_2)
= \phi_\Lambda(0.98,1.03)$, the lattice parameter $\Lambda = 0.51$, $N = 200$, and
$\sigma = \sqrt{N}$. }
\label{FG10}
\end{center}
\end{figure}

Stationary point $P_{B_2}$ is unstable in the domain of its existence $\Lambda\ge
1/3$ (except for the critical value $\Lambda=1/3$). This stationary point
corresponds to the quasi 2D stationary state, however its instability rules out
observation of 2D quantum dynamics \cite{SK}, for instance the quantum collapses
and revivals. We have found that an initial state with almost equal distribution of
atoms between two $X$-points results in the sequence of quantum recurrences, when
the wave function returns to a state with almost all atoms distributed among the
initially populated points, see Fig. \ref{FG10} (in the figure this state
corresponds to an extended population on the line $x_1 +x_2 = 1$).

\section{Discussion }
\label{conclusion}

The nonlinear  tunneling of BEC with a large number of atoms $N$ can be considered
in the semi-classical approximation, with the effective Planck constant being
$h=1/N$. We have considered the correspondence between the semi-classical regime
(equivalent to the mean-field regime) and the full quantum regime of nonlinear
tunneling between the $X$-points of the Brillouin zone of a cubic 3D lattice. In
particular, we have derived a quantum three-mode model and rewritten it as a
two-dimensional Schr\"odinger equation for an effective quantum particle, where the
effective Plank constant is $1/N$, the time scale is determined solely by the
nonlinearity of BEC, while the dynamics is controlled by a lattice parameter
$\Lambda$. The corresponding semi-classical model is the mean-field approach taking
into account the occupations of the $X$-points only. Though we have used rather
small number of atoms, $N=200$, we have found the regimes of excellent
correspondence, these are mainly about the stable stationary points of the
mean-field approach. In particular, numerical simulations show that the quantum
dynamics about the semi-classical stationary point distinguishes the stable and
unstable cases. In the case of a stable semi-classical point, one scenario consists
of  the wave function performing oscillations about the initial state with the
averages following the semi-classical dynamics. The discreteness of the quantum
energy space, however, leads to a scenario not present in the semi-classical case:
the sequence consisting of the wave function spread (i.e. becoming a nearly
phase-state) followed by the quantum recurrence to the initial nearly Fock state.
This is reflected in a deviation of the quantum averages from the semi-classical
dynamics. In the case of an unstable stationary point, the initially localized
state, i.e. nearly Fock state, is replaced by a nearly phase state  with the phases
concentrated at the semi-classical value corresponding to the unstable point (more
precisely, a linear combination of nearly phase states, since the quantum phase
appearing in the wave function of the effective quantum particle is equal to half
of the semi-classical phase due to the symmetry of the quantum Hamiltonian).

Existence of the stable stationary point with all atoms populating just one
$X$-point of the lattice allows for the experimental study of the 3D nonlinear
tunneling by modifying  the optical lattice to change the value of the lattice
parameter $\Lambda$. When the instability of the singly-populated $X$-point is
reached by varying $\Lambda$, the quantum evolution quickly establishes equal
distribution between the three degenerate  $X$-points.

\acknowledgments  The work of V.S.S. was supported by the
Visiting Professor grant from CAPES of Brazil.
The work of V.V.K. was supported by the FCT and European  program
FEDER under the grant POCI/FIS/56237/2004.

\begin{appendix}

\section{The estimate (\ref{EST})}
\label{energy}

We will use the inequality
\eqb
|\langle (b^\dag_j)^2b^2_k + (b^\dag_k)^2b^2_j \rangle|\le \langle n_j(n_j-1) +
n_k(n_k-1)\rangle,
\label{H1}\eqe
which follows from
\begin{eqnarray*}
&& 0\le\langle\Phi|(A +B)^\dag(A+B)|\Phi\rangle
\\
&&= \langle\Phi|A^\dagger A|\Phi\rangle + \langle\Phi|B^\dagger B|\Phi\rangle +
\langle\Phi|A^\dagger B|\Phi\rangle+\langle\Phi|B^\dagger A|\Phi\rangle
\end{eqnarray*}
by setting $A= b_j^2$ and $B=\pm b_k^2$. Using (\ref{H1})  we obtain
\begin{eqnarray}
 \langle\hat{H}_-\rangle\le \langle \Hat{H}\rangle \le \langle\hat{H}_+\rangle,
\label{H2}
\end{eqnarray}
where $\hat{H}_{\pm}$ are two $c$-number operators:
\begin{eqnarray*}
\hat{H}_{\pm} = \frac{1}{4}\sum_{j=1}^3\frac{n_j^2}{N^2}
+\Lambda\sum_{j<k}\frac{n_j}{N}\frac{n_k}{N}
\pm\frac{\Lambda}{2}\sum_{j=1}^3\frac{n_j}{N}\frac{n_j-1}{N},
\label{H3}
\end{eqnarray*}
which may be treated as classical functions of $n_j$. Next, reducing the total
squares $(\sum_{j=1}^3n_j/N)^2 = 1$ in $\hat{H}_\pm$ we have
 \begin{eqnarray}
\label{H+}
     \hat{H}_+ =\frac14\sum_{j=1}^{3} \frac{n_j^2}{N^2}
    +\frac{\Lambda}{2}\left(1-\frac 1N\right),
\\
\label{H-}
    \hat{H}_-=\frac{1-4\Lambda}{4}\sum_{j=1}^3\frac{n_j^2}{N^2}
    +\frac{\Lambda}{2}\left(1+\frac{1}{N}\right).
\end{eqnarray}
The inequalities (\ref{EST}) follow from   (\ref{H+}) and (\ref{H-}) if one takes
into account that $N^2/3\le \sum_{j=1}^{3} {n_j^2} \leq N^2$.

\section{The semi-classical stationary solutions with only two
populated X-points}

Consider the Hamiltonian (\ref{Heff}) in the semi-classical limit, i.e. $b_j\to
\sqrt{N}b^{(cl)}_j$. Suppose that there is a stationary point with $b_{10} =
\cos\theta e^{-i\phi_1/2}$, $b_{20} = \sin\theta e^{-i\phi_2/2}$, and $b_{30} = 0$
with some $0<\theta<\pi/2$. To find all such stationary points we use that the
semi-classical Hamiltonian expanded about such a point does not have any linear
terms. Without loss of generality we can set
\begin{eqnarray}
b_1  &=& \cos\theta e^{-i\phi_1/2}(1 + \beta_1),\, b_2 = \sin\theta
e^{-i\phi_1/2}(1
+\beta_2),\nonumber\\
b_3 &=& \beta_3 \in \mathrm{Re}
\end{eqnarray}
where, due to the conservation of the
number of atoms, in the linear order we get
\eqb
\beta_3^2 = - \cos^2\theta(\beta_1 +\beta_1^*) - \sin^2\theta(\beta_2
+\beta_2^*).
\eqe
Using this we obtain the linear in $\beta_{1,2}$ terms as follows:
\begin{eqnarray}
&&\frac{\partial\mathcal{H}}{\partial \beta_1}\biggr|_{\beta_{1,2}=0}
=\cos^2\theta\biggl[\frac{\cos^2\theta}{2} -\Lambda\cos^2\theta
\nonumber\\
&&+ \frac{\Lambda}{2}\biggl(\sin^2\theta e^{i(\phi_2-\phi_1)}
-
\cos^2\theta\cos\phi_1
-\sin^2\theta\cos\phi_2\biggr) \biggr].
\nonumber\\
\label{A3}
\end{eqnarray}
In calculation of (\ref{A3}) we have used that the contributing terms are
\begin{eqnarray*}
\mathcal{H}_\mathrm{\beta_1} &=& \frac{1}{4}n_1^4 +\Lambda[n_1n_2+ n_1n_3+ n_2n_3]
\\
&+&\frac{\Lambda}{4}[(b_2^*)^2b_1^2 + (b_3^*)^2b_1^2 +(b_3^*)^2b_2^2
+(b^*_2)^2b_3^2].
\end{eqnarray*}
By changing $\cos\theta\to\sin\theta$ and $\phi_1\to\phi_2$ we have also
\begin{eqnarray}
&&\frac{\partial\mathcal{H}}{\partial \beta^*_2}\biggr|_{\beta_{1,2}=0}
=\sin^2\theta\biggl[\frac{\sin^2\theta}{2} -\Lambda\sin^2\theta
\nonumber\\
&&+ \frac{\Lambda}{2}\biggl(\cos^2\theta e^{i(\phi_2-\phi_1)}-
\sin^2\theta\cos\phi_2
-\cos^2\theta\cos\phi_1\biggr) \biggr].
\nonumber\\
\label{A4}\end{eqnarray}
The r.h.s.'s in equations (\ref{A3}), (\ref{A4}) and in their complex conjugates
should give zero for a stationary solution. First of all, from (\ref{A3}) we have
$\sin(\phi_2-\phi_1)=0$, hence $\phi_2 = \phi_1$ or $\phi_2 = \phi_1 \pm\pi$. Then,
combining  equations (\ref{A3}), (\ref{A4}) we get
\[
(\cos^2\theta-\sin^2\theta)\left(\frac12 -\Lambda
-\frac{\Lambda}{2}\cos(\phi_2-\phi_1)\right)=0,
\]
i.e. ($i$) $\cos\theta = \sin\theta = 1/\sqrt{2}$ or ($ii$)
$\cos(\phi_2-\phi_1)=\frac{1-2\Lambda}{\Lambda}$. In case ($i$) we obtain the phase
$
\cos(\phi_\Lambda) = \frac{1-\Lambda}{2\Lambda},
$
while in case ($ii$) there is no solution except for the special value of the
lattice parameter $\Lambda = 1/3$ when the phase $\phi_1$ becomes arbitrary. Hence,
we arrive at the stationary point
$
(x_1,x_2,x_3)=\left(\frac12,\frac12,0\right)$,
$\phi_1 = \phi_2 = \phi_\Lambda,$
which exists only for $\Lambda\ge\Lambda_1 = 1/3$ (the phases appear in the initial
state (\ref{INCOND})). In terms of $z$-variables the stationary point reads $z_1 =
z_2 = 0$.
\end{appendix}

\end{document}